\documentclass[prd,notitlepage,showpacs,preprintnumbers,amsmath,amssymb,nofootinbib,10pt,superscriptaddress,twocolumn]{revtex4-2}

\usepackage{dcolumn}
\usepackage{bm}
\usepackage{xcolor}
\usepackage[utf8]{inputenc}
\usepackage[spanish,english]{babel}
\usepackage{amsmath,amssymb,amsfonts,latexsym,cancel}
\usepackage{graphicx}
\usepackage{color}
\usepackage{soul}
\usepackage{ulem}
\usepackage{hyperref}
\usepackage{amsmath}
\usepackage{slashed}
\usepackage{braket}
\usepackage{amssymb}
\usepackage{amsmath}
\usepackage{float}


\newcommand{\be}{\begin{equation}}
    \newcommand{\ee}{\end{equation}}
\newcommand{\bea}{\begin{eqnarray}}
    \newcommand{\eea}{\end{eqnarray}}

\newcommand{\dx}{{\rm dx}}

\newcommand{\CL}{{\tt ${\mathcal C}$osmo${\mathcal L}$attice}~}
\newcommand{\CLns}{{\tt ${\mathcal C}$osmo${\mathcal L}$attice}}

\begin{document}

\title{The Spectrum of Gravitational Waves from Annihilating Domain Walls}

\newcommand{\addressICCUB}{Departament de Física Quàntica i Astrofísıca \& Institut de Ciències del Cosmos (ICCUB),
    Universitat de Barcelona, Martí i Franquès 1, 08028 Barcelona, Spain}
    
\author{Alessio Notari}
\affiliation{\addressICCUB}
\author{Fabrizio Rompineve}
\affiliation{Departament de F\'isica, Universitat Aut\`onoma de Barcelona, 08193 Bellaterra, Barcelona, Spain
\looseness=-1}
\affiliation{Institut de F\'isica d’Altes Energies (IFAE) and The Barcelona Institute of Science and Technology (BIST), Campus UAB, 08193 Bellaterra (Barcelona), Spain}
\author{Francisco Torrenti}
\affiliation{\addressICCUB}

\date{\today}

\begin{abstract} 
Networks of cosmic domain walls can form in the early Universe as a consequence of the spontaneous breaking of discrete symmetries.
We study the production of a cosmological background of gravitational waves (GWs) from such networks, when they annihilate due to a small explicit symmetry breaking term. Averaging over several 3+1-dimensional high-resolution lattice field simulations, we obtain a GW spectrum with the following characteristics: (1) a broad asymmetric peak, roughly located at frequency (at the time of emission) $f\sim 2 H_{\rm gw}$, where $H_{\rm gw}$ is the Hubble rate at the end of GW production, shortly after annihilation, (2) a doubly broken power law spectrum $\propto k^{-n}$, with initial slope $n \sim 0.5$ after the main peak and $n \sim 1.8$ at high $f$, while the low frequency region $f<f_p$ agrees with the causality behavior $\sim k^3$. Additionally, extending previous results, we find that GW production continues to be efficient until a value of the Hubble scale $H_{\text gw}$ that is roughly an order of magnitude smaller than the naive estimate $\sigma H = \Delta V$, where $\sigma$ is the wall tension and $\Delta V$ the size of the symmetry breaking term, thereby leading to a $O(100)$ larger GW signal. 
We find such results to be robust when changing the shape of the scalar field potential or including a time-dependent symmetry breaking term. Our findings have important implications for GW searches, especially in light of the reported evidence for a stochastic GW background in Pulsar Timing Array data.
 \end{abstract}

\maketitle

\section{Introduction}

Topological defects in field theories have been investigated as possible components of the early Universe since the early times of modern cosmology~\cite{Zeldovich:1974uw, Kibble:1976sj} (see also~\cite{Vilenkin:2000jqa} and references therein for a comprehensive introduction). Their existence would be a consequence of patterns of symmetry breaking that arise in several well-motivated scenarios of UV physics beyond the Standard Model (SM), from Grand Unification Theories (GUTs)~\cite{Preskill:1979zi} to QCD axion models~\cite{Sikivie:1982qv}, to scalar extensions of the SM, that may also arise from naturalness motivated scenarios, e.g.~\cite{Espinosa:2011eu, Arkani-Hamed:2020yna}. Due to the typically high energy scale associated to the specific UV physics, and additionally to their peculiar cosmological evolution, topological defects are often regarded as dangerous species that may overclose the Universe (as is the case of GUT monopoles~\cite{Zeldovich:1978wj, Preskill:1979zi}) if they are produced after inflation. Such a potentially important impact on observations however also represents an opportunity to probe their existence and the microphysics that generates them, when viable scenarios can be constructed.

The two-dimensional case, i.e.~domain walls (DWs), is realized when a discrete symmetry is spontaneously broken in the cosmological evolution, and arguably provides the best example of such an opportunity. Indeed, DWs stand out compared to defects of lower dimensionality because of two reasons. First, they arise already in the simple field theory of a real scalar field with a $\mathbb{Z}_2$-symmetric potential (in contrast to monopoles and strings, which require more degrees of freedom). Second, while their evolution achieves a scaling regime which closely resembles that of cosmic strings~\cite{Zeldovich:1974uw, Kibble:1976sj}, whereby an approximately fixed number of Hubble-sized defects is present at each epoch in the cosmological evolution, their higher spatial dimensionality implies that their energy density redshifts slower than radiation and matter, i.e. $\rho_\text{dw}\sim \sigma H\sim a^{-2}$, where $\sigma$ is the domain wall mass per unit surface (tension), $H$ the Hubble expansion rate and $a$ is the scale factor. Such a slow dilution rules out the possibility of spontaneously broken, but otherwise exact, discrete symmetries in the radiation epoch~\cite{Zeldovich:1974uw}, if the resulting DW tension is larger than about the $\text{MeV}^3$ scale. Nonetheless, as long as a sufficiently large explicit breaking of the discrete symmetry is present~\cite{Sikivie:1982qv}, that may be expected on general grounds, domain walls can decay in the early Universe, while they are still subdominant compared to the radiation background (see e.g.~\cite{Vilenkin:1981zs, Coulson:1995nv, Larsson:1996sp, Ramazanov:2021eya,Gonzalez:2022mcx,Blasi:2022woz,Babichev:2023pbf} for other possibilities to evade the so-called domain wall problem). 

Domain walls are peculiar in yet another aspect, that is of core relevance for their possible detection. Since they have no additional light degrees of freedom to radiate into, while being also characterized by large quadrupolar inhomogeneities and relativistic speed, they are very efficient sources of cosmological gravitational waves (GWs) \cite{Vilenkin:1981zs, Preskill:1991kd, Chang:1998tb, Gleiser:1998na, Hiramatsu:2013qaa} (gauge cosmic strings can also be similarly strong sources, see \cite{Wachter:2024zly} for recent progress). Previous studies~\cite{Hiramatsu:2013qaa,Ferreira:2024eru, Dankovsky:2024zvs} have indeed found that the energy density in GWs radiated by DWs almost saturates the simple dimensional analysis prediction from the quadrupole formula, i.e. $\Omega_\text{gw}\equiv \rho_\text{gw}/(3H^2 M_p^2)\simeq 3/(32\pi)\alpha_\text{dw}^2$, where $\alpha_\text{dw}\equiv \rho_\text{dw}/(3H^2 M_p^2)$ is the energy density fraction of the Universe in domain walls at any value of $H$. Since $\alpha_\text{dw}\propto a^2$ in the radiation epoch until DW annihilation occurs, the GW signal from domain walls is dominated by the very latest stages of their evolution. Thus, while they are in principle long-lived sources, their GW spectrum is peaked at about the frequency corresponding to the Hubble scale at the time of their annihilation. The current sensitivities of ground based interferometers (LIGO/Virgo/KAGRA, or LVK) and Pulsar Timing Arrays (PTAs) allow to probe the presence of domain walls making a fraction $\alpha_\text{dw}\gtrsim 0.05$ of the Universe at temperatures either $10^7~\text{GeV}\lesssim T\lesssim 10^9\text{GeV}$ (LVK) or $\text{MeV}\lesssim T\lesssim \text{GeV}$ (PTAs). Future 3G and space-based interferometers should be able to push the exploration down to $\alpha_\text{dw}\gtrsim 0.001$ across a much wider temperature range (see~\cite{Caprini:2024ofd} for a recent assessment). 

In light of such an exciting observational outlook, it becomes important to obtain a robust and detailed characterization of the GW signal from domain walls, which goes beyond simple analytical estimates and numerical approximations. This is the aim of this work. Such an effort is made particularly timely by the recently reported evidence for a stochastic GW background at nHz frequencies by PTA collaborations~\cite{NG15-SGWB, EPTA2-SGWB, PPTA3-SGWB, CPTA-SGWB}, with domain walls being among the non-astrophysical sources that provide a good interpretation of the data~\cite{Ferreira:2022zzo, NANOGrav:2023hvm}. With upcoming data releases from PTAs, detailed spectra will be necessary to discriminate between an early Universe origin of the source and the astrophysical interpretation, or to set constraints on new physics.

Due to the non-linearities that are intrinsic  to topological defects, a detailed characterization of the GW signal from domain walls requires numerical (lattice classical field theory \cite{Figueroa:2020rrl}) simulations of domain wall formation, evolution and annihilation in the expanding Universe. A similar necessity has long been realized and addressed for other primordial GW sources, in particular first order phase transitions and local cosmic strings, see e.g.~\cite{LISACosmologyWorkingGroup:2022jok} for a recent collection of results. The first simulations with this aim were performed more than a decade ago~\cite{Hiramatsu:2013qaa}, also for hybrid string-wall networks~\cite{Hiramatsu:2012sc}, focusing on the scaling regime (see also~\cite{Kawasaki:2014sqa, Correia:2014kqa,Correia:2018tty} for DW simulations without GWs). The results of these works have since then been used both for GW searches (in LVK~\cite{Jiang:2022svq} and PTA data~\cite{Ferreira:2022zzo,NANOGrav:2023hvm}) as well as for phenomenological analyses of particle physics models. Very recently, updated numerical studies have appeared, focused on: specific axion models for DW annihilation~\cite{Kitajima:2023cek}; dark matter production from the decay of domain walls~\cite{Chang:2023rll}; the estimation of the fraction of Primordial Black Holes (PBHs) that can be produced by DW collapse~\cite{Ferreira:2024eru} (see also~\cite{Pujolas:2022qvs}); the GW signal from the scaling regime, both in matter~\cite{Ferreira:2023jbu} and radiation-domination~\cite{Kitajima:2023kzu, Dankovsky:2024zvs}, i.e.~not including the annihilation phase, and the GW signal from melting domain walls \cite{Dankovsky:2024ipq}.

In this work, we aim to improve on those previous works, with a focus on the GW spectrum from domain wall annihilation. This has been previously shown to be the dominant contribution to the GW signal in a realistic DW cosmology~\cite{Kitajima:2023cek, Ferreira:2024eru}. It is thus this spectrum that should be used for searches in GW datasets, rather than the previously employed spectrum from the scaling regime (which we will update as well en route to our main results). Besides the focus on the annihilation phase, our novelties compared to the aforementioned works are several: first, we employ improved numerical techniques and computational capabilities that allow to isolate the impact of numerical artifacts in the high frequency part of the spectrum; second, we consider several classes of DW models, with different well-motivated possibilities for the scalar field potential and the explicit symmetry breaking term that induces DW annihilation; third, we present analytical fits to the numerical results for the GW spectra, that can be readily used to perform searches for cosmic DWs in GW datasets (LVK, PTAs as well as future data).

Our paper is structured as follows. In Sec.~\ref{sec:scaling} we introduce the basic equations and  numerical techniques used in this work. In Sec.~\ref{sec:biased} we present the main results of this work: the GW spectrum from an annihilating domain wall network with a time-independent explicit symmetry breaking term and a $\mathbb{Z}_2$-symmetric quartic potential. In Sec.~\ref{sec:variants} we explore variations of this main model, in particular employing a periodic potential and a time-dependent symmetry breaking term. We summarize and discuss our results in Sec.~\ref{sec:conclusions}. Our work also comes with two Appendices: in App.~\ref{sec:unbiased} we present complementary simulations for domain walls in the absence of a symmetry breaking term, while in App.~\ref{app:technicalities} we provide additional material on our lattice results.

\section{Scaling regime and outline of lattice simulations}
\label{sec:scaling}

We shall start with a simple model capable of sustaining domain walls (DWs): a scalar field $\phi$ with a $\mathbb{Z}_2$ symmetric potential of the form 
\be V(\phi) = \frac{\lambda}{4} (\phi^2 - v^2)^2 \ . \label{eq:quartic-pot}\ee
The potential has two degenerate minima at the field values $\phi = \pm v$, and the mass around them is $m = \sqrt{2 \lambda} v$. We take the field to be initially mostly homogeneous, with initial value $\phi_i\approx 0$, as is appropriate when high temperature effects in the early Universe induce a temperature-dependent quadratic potential that drives the field to the minimum. Due to random fluctuations, the field starts rolling towards either of both minima roughly once the Hubble parameter $H$ is equal to the mass $m$, hence forming a DW network.  Shortly after its formation, the network attains a `scaling regime' characterized by the existence on average of a fixed $O(1)$ number of DWs per Hubble patch. The physical width of the domain walls is approximately $\delta_w \equiv m^{-1}$, and their tension (i.e.~the energy per unit area) is $\sigma \equiv \int_{-v}^{+v}\sqrt{2V(\phi)}d\phi=  2 \sqrt{2 \lambda} v^3/3 $.
In the expanding Universe with FLRW metric with scale factor $a$, conformal time $\eta$,  and $H\equiv a' /a^2\equiv \mathcal{H}/a $ the Hubble rate ($'\equiv d/d\eta$ and $\mathcal{H}$ is the conformal Hubble rate), the energy density of the DW network can hence be written as
\be \rho_{\rm dw} = \frac{\sigma A a^2}{V a^3} = 2 \mathcal{A} \sigma H \ , \hspace{0.5cm} \mathcal{A} \equiv \frac{A}{V} \frac{1}{2 a H} \ , \label{eq:area-parameter} \ee
where $A$ is the total comoving area of the DWs in a box of comoving volume $V$ and $\mathcal{A}$ is the so-called \textit{area parameter}, which remains approximately constant and of order unity in the scaling regime. In this work we focus on a radiation-dominated Universe with $a \propto \eta$, so that the energy density of the DW network evolves as $\rho_{\rm dw} \propto H \propto \eta^{-2}$ in the scaling regime.

Our work focuses on the stochastic gravitational wave background (SGWB) generated by the network of domain walls. The energy density of such GWs is (see e.g.~\cite{Maggiore:2018sht})
\be \rho_{\rm gw} \equiv \frac{M_p^2}{4 a^2} \langle h'_{ij} h'_{ij} \rangle_V \ , \ee
where $h_{ij}$ is the stochastic GW field (in the tranverse-traceless, TT, gauge) and $\langle \dots \rangle_V$ is an average over a comoving volume large enough to capture all the relevant wavelengths. We can write this expression in momentum space as follows,
\begin{align}  \rho_{\rm gw}& = \int \frac{d \rho_{\rm gw}}{d \log k} \, d \log k \ , \\
 \frac{d \rho_{\rm gw}}{d \log k} & =  \frac{m_p^2 k^3}{8 \pi^2 a^2 V} \int \frac{d \Omega_k}{4 \pi} h'_{ij} ({\bf k}, \eta) h'^*_{ij}  ({\bf k}, \eta)  \ , \label{eq:RhoGWlogk}
\end{align}
where $d \rho_{\rm gw} / d \log k$ is the energy density per logarithmic momentum interval and $d\Omega_k$ the solid angle in momentum space.  It is customary to define $\Omega_{\rm gw} \equiv \rho_{\rm gw} / \rho_c $ as the fraction of the critical energy density $\rho_c \equiv 3 M_p^2 H^2$ in GWs at a given epoch, with $M_p=1/\sqrt{8\pi G}\simeq 2.4\cdot 10^{18}~\text{GeV}$. A simple dimensional analysis estimate by means of the quadrupole formula reveals that $\rho_{\rm gw}$ should be quadratic in the energy density of the source and suppressed by the gravitational coupling, leading to the generic expectation $\rho_{\rm gw}\approx \sigma^2/M_{p}^2$, which is constant in time,\footnote{Since $\rho_\text{gw}\approx \text{const}$ in the scaling regime, while $\rho_\text{dw}\propto a^{-2}$, the simple quadrupole estimate above makes sense only as long as $\rho_\text{gw}^{\text{quad}}\ll \rho_\text{dw}$. It is straightforward to see that this condition is respected as long as DWs do not dominate over the radiation background. Afterwards, the scaling regime would break down anyway and the estimates above would not be valid. This condition is anyway required for phenomenological viability of the scenario.} and hence $\Omega_{\rm gw} \sim \eta^2$.

Previous numerical studies of DW networks in the scaling regime have indeed confirmed the validity of this estimate for the amplitude of the GW spectrum at the peak frequency~\cite{Hiramatsu:2013qaa, Ferreira:2024eru, Dankovsky:2024zvs}. This can also be written in a convenient form that is valid for a general horizon-sized relativistic causal GW source (i.e. mostly active on sub-horizon scales at a given epoch), see e.g.~\cite{Saikawa:2017hiv}:
\be 
\label{eq:quadrupole}
\Omega_{\rm gw}^{\rm (scal)} (k_{\rm peak}, \eta ) = \epsilon~\Omega_\text{gw}^\text{quad}=\frac{3}{32 \pi} \epsilon \,  \alpha (\eta)^2 , \ee
with $\alpha(\eta) \equiv \rho_{\rm source}/\rho_c$ the fraction of the universe's energy density stored in the GW source, and $\epsilon \simeq \mathcal{O} (1)$ an efficiency factor that can be determined with numerical simulations. For the case of DWs in the scaling regime, $\rho_\text{source}=\rho_{\rm dw} \simeq 2 \mathcal{A} \sigma H$.

Let us now provide an outline of the numerical techniques and strategy adopted in this work (we refer the interested reader to App.~\ref{app:technicalities} for further details). In our work we employ a modified version of the publicly-available code \CL \cite{Figueroa:2021yhd}, which solves a discretized version of the following equations of motion,
\begin{align}
    \phi'' - \nabla^2 \phi + 2 \mathcal{H} \phi'&=  - a^2 \partial_{\phi} V  \ , \label{eq:eomscalar} \\
    h''_{ij} - \nabla^2 h_{ij} + 2 \mathcal{H} h'_{ij} &= \frac{2}{m_p^2} (\partial_i \phi \partial_j \phi )^{\rm TT} \ ,  \label{eq:eomGW} 
\end{align}
where $\phi = \phi (\vec{x}, t)$ is the scalar field and $h_{ij} = h_{ij}  (\vec{x},t)$ is sourced by the TT part of the scalar field anisotropic tensor. Throughout our work we assume that the DWs contribute only a subdominant fraction of the universe energy density and thus fix the background to be radiation dominated at all times. As initial conditions we set the homogeneous component of the scalar field to $\bar{\phi} = 0$, over which we impose a white spectrum of small fluctuations. The equations of motion \eqref{eq:eomscalar}-\eqref{eq:eomGW} are solved in 3-dimensional regular lattices of $N$ points per dimension and comoving side length $L$ with periodic boundary conditions. The minimum and maximum comoving momenta captured by such lattice are $k_{\rm IR} \equiv 2 \pi /L$ and $k_{\rm UV} \equiv  \sqrt{3} \pi /\Delta x$ respectively, with $\Delta x \equiv L/N$ the lattice spacing \cite{Figueroa:2020rrl}. We refer the reader to \cite{GWTechNote} for a detailed explanation on how $\rho_\text{gw}$ is extracted from the lattice simulations.

In order to trust the results of our simulations, $N$ and $L$ must be chosen appropriately so that we correctly capture two relevant scales within our lattice. First, the comoving width of the wall decreases as the universe expands, so our lattice must have enough resolution at small scales to properly resolve it until the end of the simulation. Second, the number of Hubble patches contained in the lattice decreases with time as $c (\eta) \equiv \left( \frac{a (\eta) L}{H^{-1} (\eta)} \right)^3 \propto \eta^{-3}$, and in the scaling regime we expect approximately one DW per Hubble patch, so we also require $c > 1$ at the end of the simulation. The comoving momenta scales associated to the wall width and to the Hubble patch are respectively,
\be k_{\text{w}} = \frac{2 \pi a (\eta) }{\delta_w} \propto \eta \ , \hspace{0.5cm}  k_{\text{h}} = \frac{2 \pi a(\eta)}{H^{-1} (\eta)} \propto \eta^{-1} \ , \label{eq:momscales} \ee
and hence their separation grows with time. Therefore, in order to properly simulate the DW dynamics, we must choose $N$ and $L$ such that $k_{\rm IR} < k_{\text{h}} < k_{\text{w}}  < k_{\rm UV}$ at all times. From these conditions, given a value of $N$, the maximum time that can be simulated $\eta_{\rm max}$ and the number of Hubble patches within the lattice at that time are
\begin{align}
\eta  < \eta_{\rm max} &= \left(\frac{\sqrt{3}}{2} \frac{N}{H_i L}  - 1 \right) H_i^{-1} \ , \label{eq:tmax} \\
c(\eta_{\rm max}) & =\left( \frac{2 L^2 H_i^2}{\sqrt{3} N}\right)^3 \ , \label{eq:cmax}
\end{align}
where we have assumed a radiation-dominated expansion $a(\eta) = 1 + H_i \eta$, with $H_i \equiv H(\eta = 0)$ the Hubble parameter at the onset of the simulation. Note that the lattice length must obey $L \geq L_{\rm min} \equiv (3^{1/4} / H_i) \sqrt{N/2}$ in order to capture at least one Hubble patch within the lattice at the time $\eta_{\rm max}$.

Accurate simulations of domain walls require large lattices of $N \gtrsim 10^3$ points per dimension, which is computationally very expensive. In order to improve the reliability of our results and minimize the computation costs, in our simulations we have employed two numerical ingredients that are novel compared to previous literature on the topic. First, we have accelerated the emergence of the scaling regime by adding a friction term to the LHS side of the scalar field EOM \eqref{eq:eomscalar} during some time interval before the formation of the DWs. This term is chosen so that the field relaxes to the minima with negligible oscillations, converging much quicker to the scaling solution. Details about the implementation of friction are given in App.~\ref{app:spatialderivs}. There we also compare simulations with and without friction, and show that the final values for the DW network area parameter and energy components coincide. 

Second, in our simulations we approximated the spatial derivatives appearing in the EOM \eqref{eq:eomscalar}-\eqref{eq:eomGW} by discretized versions accurate up to fourth order in the lattice spacing $\Delta x$, instead of the standard second-order approximations used by e.g.~the public version of \CL \cite{DiscreteDerivatives}. These allow to improve the accuracy of the simulations at intermediate momenta scales. In App. \ref{app:technicalities} we show an explicit comparison between simulations using second- and fourth-order spatial derivatives. All the results presented in this work incorporate friction and fourth-order derivatives unless otherwise stated. These novelties allow to better resolve the shape of the GW spectrum without the need of increasing the number of lattice points.

Before moving on to the main scenario of interest, let us describe the units used in our numerical analysis. We set $v=1$ and $m=\sqrt{2 \lambda} v = 1$ to simplify notation. In these units we have $\lambda = 1/2$, and $\sigma = 2/3$, and both conformal time $\eta$ and the lattice side length $L$ are dimensionless. We also fix $a(\eta=0)=1, H_i=H(\eta=0)=1$, such that the evolution of the scale factor is $a (\eta) = 1 + \eta$.

\begin{figure}
    \begin{center}
        \includegraphics[width=0.46\textwidth]{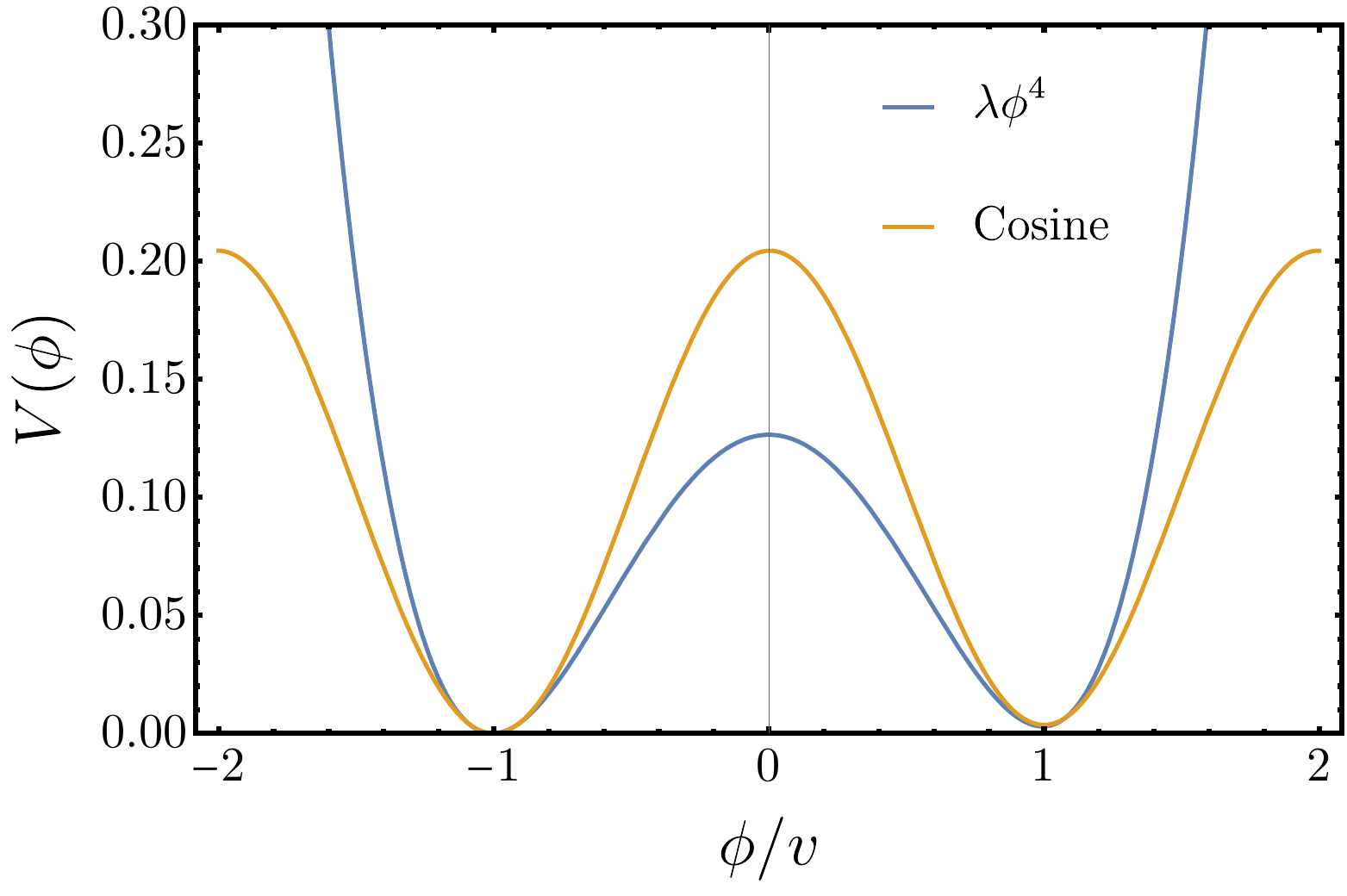}
    \end{center} \vspace{-0.4cm}
    \caption{Potentials considered in this work.}
\label{fig:potentials}
\end{figure}

\section{Main model: $\lambda\phi^4$ with T-independent bias}
\label{sec:biased}

So far we have been describing the scenario where the two minima are exactly degenerate and the network achieves a scaling regime which lasts until DWs dominate over the background radiation. However, a cosmologically viable and abundant domain wall network necessarily requires an annihilation mechanism whereby the scaling regime ends before the network comes to dominate the Universe, and the walls disintegrate into non-relativistic or mildly relativistic scalar particles.

\begin{figure*}[t] 
    \begin{center}
        \includegraphics[width=0.47\textwidth]{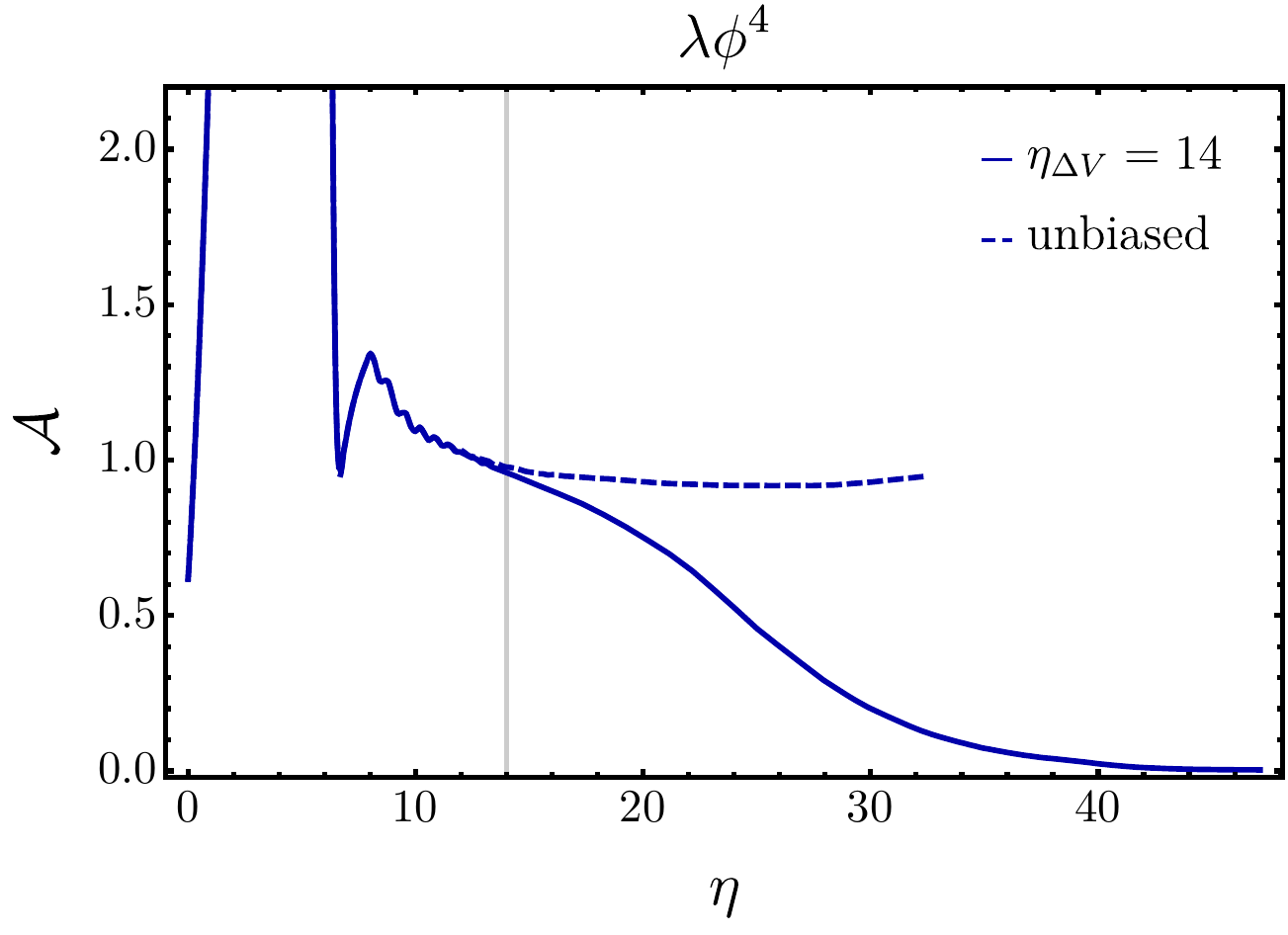}
        \hspace{1em}
        \includegraphics[width=0.46\textwidth]{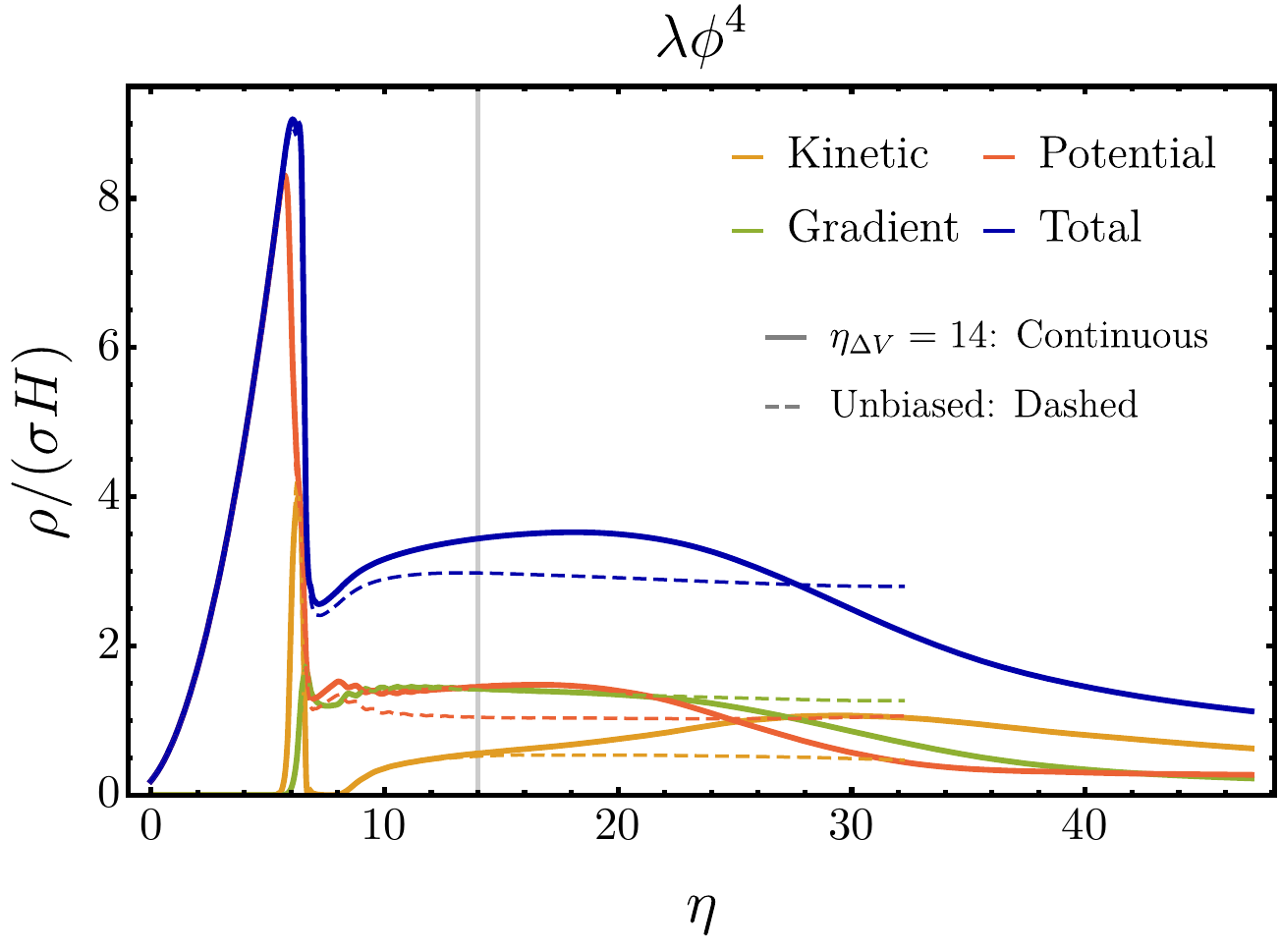}
    \end{center} \vspace{-0.3cm}
    \caption{{\it Left:} Evolution of the area parameter in the presence of a symmetry breaking term in the potential (solid curve), obtained from a simulation with $N=3060$ and $L=80$. The size of $\Delta V$ has been chosen such that $\sigma H(\eta_{\Delta V} )=\Delta V$ at $\eta_{\Delta V}=14$ (vertical gray line). The evolution of the area parameter in the absence of the symmetry breaking term is also shown for comparison (dashed curve). {\it Right:} Evolution of the energy density components of the scalar field, for a biased (solid) and unbiased (dashed) network.}
\label{fig:areaparam}
\end{figure*}

\begin{figure*}[t] 
    \begin{center}
        \includegraphics[width=0.3\textwidth]{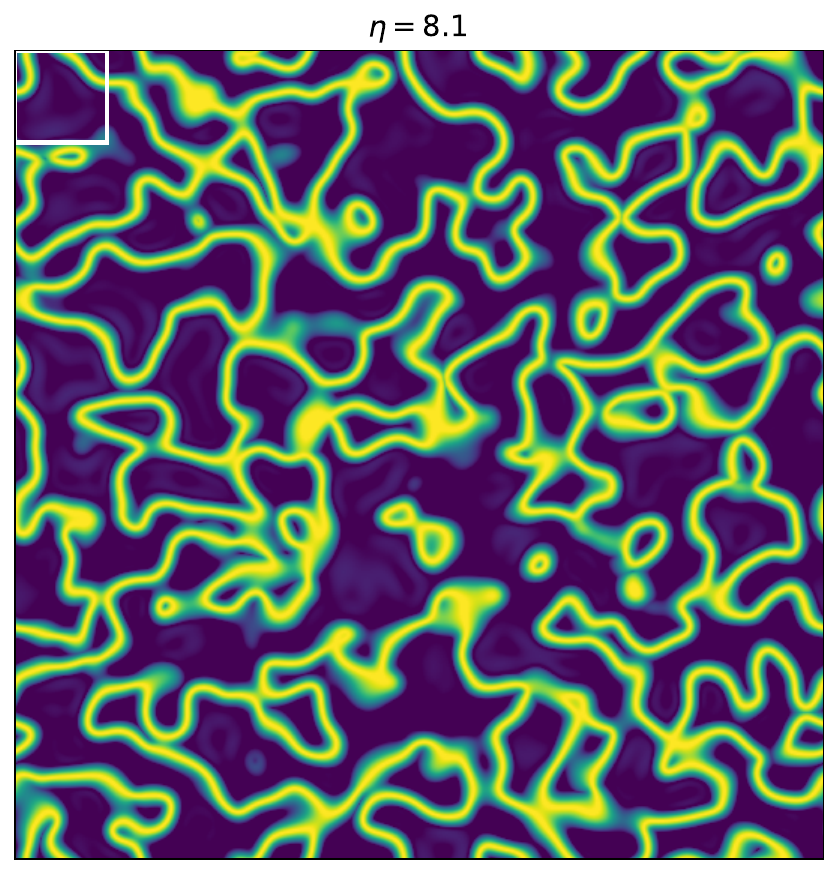}\hspace{1em}
        \includegraphics[width=0.3\textwidth]{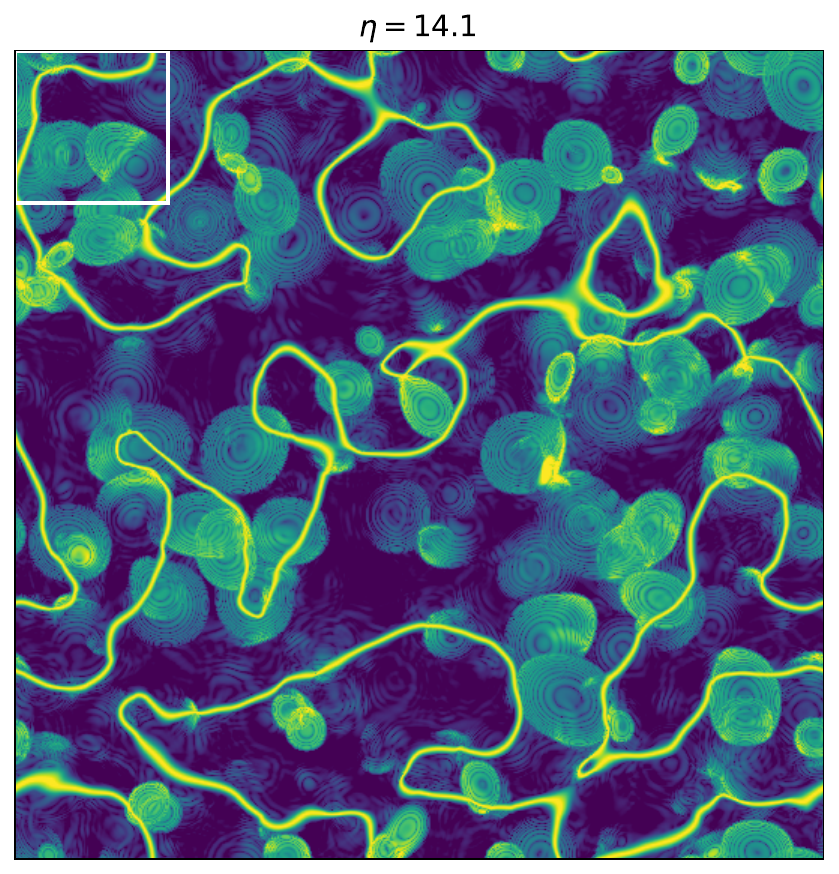}\hspace{1em} 
        \includegraphics[width=0.3\textwidth]{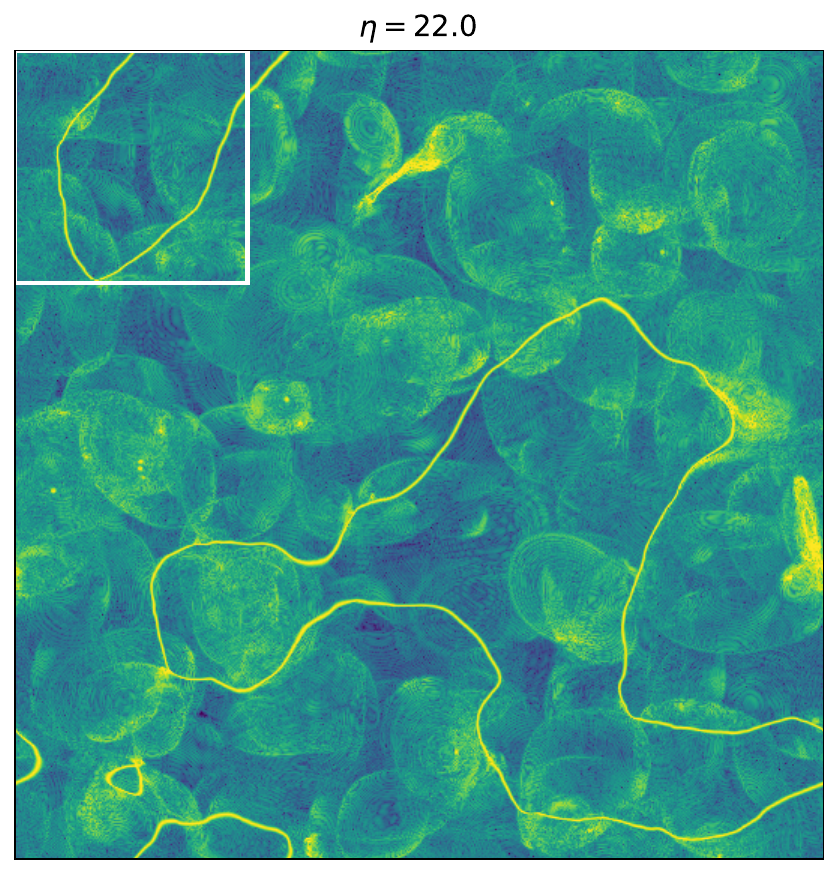}\\
        \includegraphics[width=0.3\textwidth]{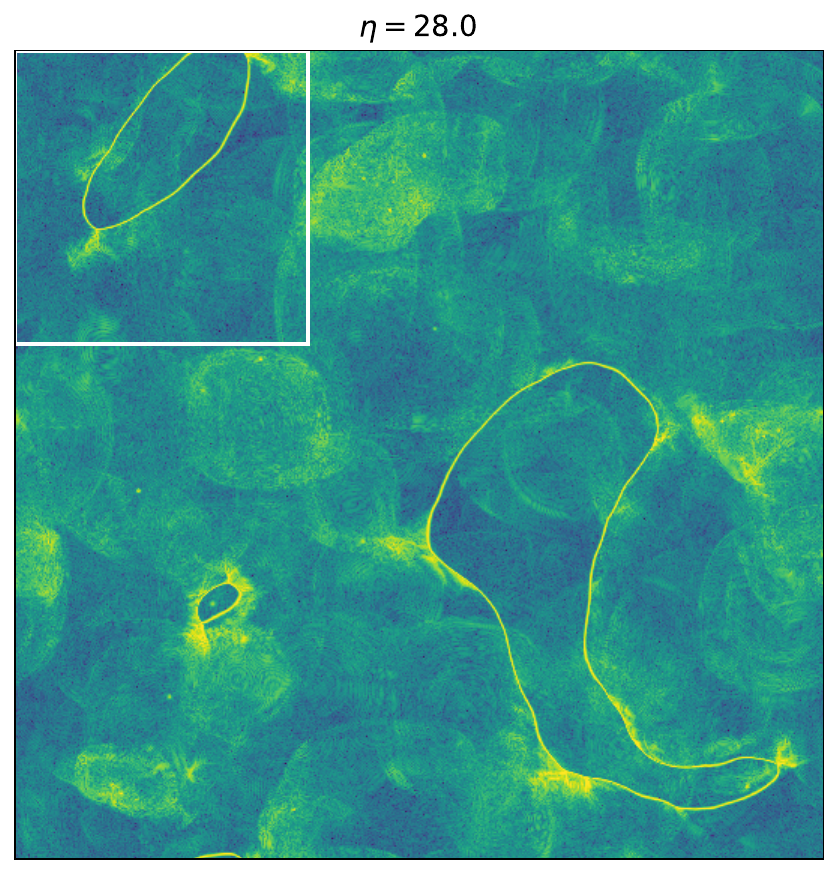}\hspace{1em}
        \includegraphics[width=0.3\textwidth]{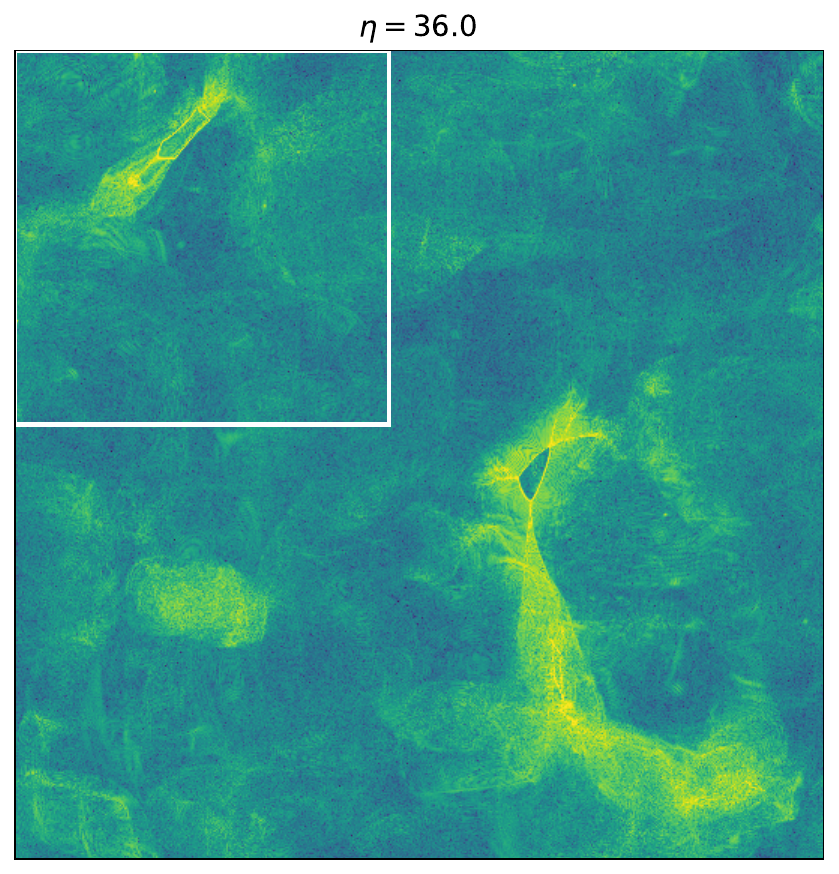}\hspace{1em} 
        \includegraphics[width=0.3\textwidth]{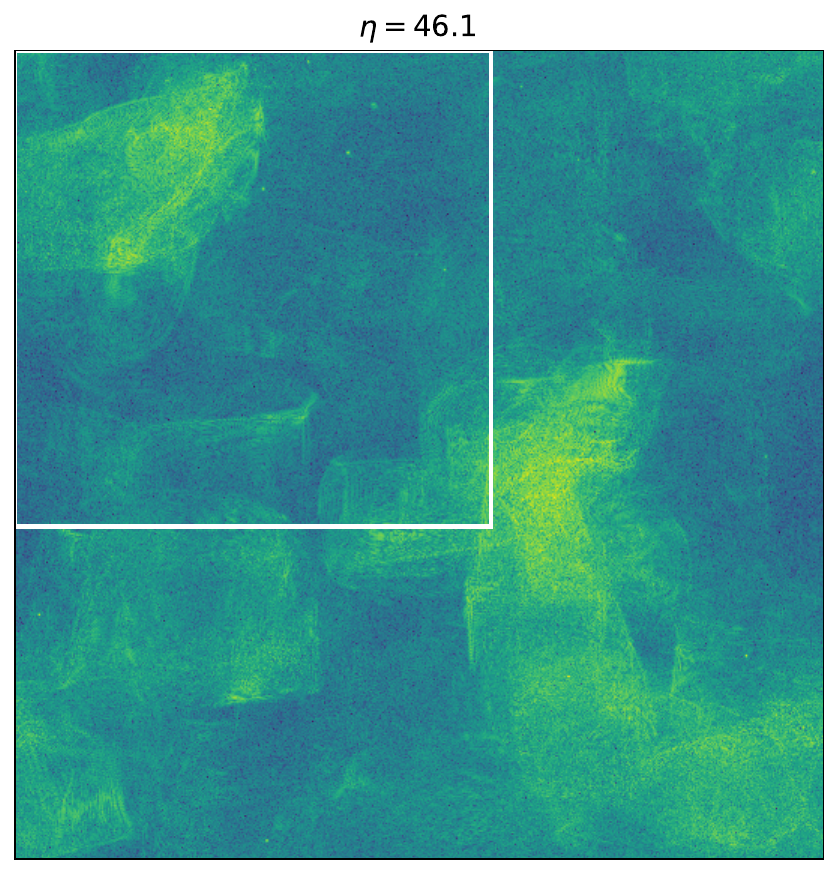}\\ \vspace{0.4cm}
        \includegraphics[width=0.5\textwidth]{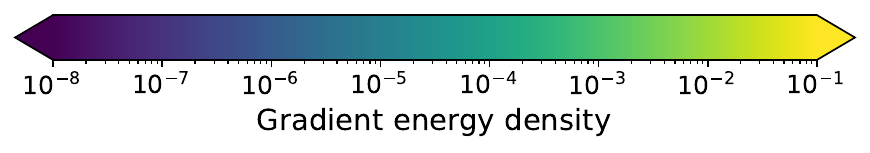}
    \end{center}
    \caption{2D snapshots of the gradient energy density (in logarithmic scale) for the biased $\phi^4$ potential at different times, extracted from a simulation with $N=3060$ and $L=80$. The darkest and lightest colors represent, respectively, gradient energy densities smaller than $10^{-8}$ and larger than $10^{-1}$, in units of $m=v=1$. The squares in each panel have sides of length $1/H$. }
\label{fig:snapshots}
\end{figure*}

In this work, we consider a simple annihilation mechanism: we add a small $\mathbb{Z}_2$-breaking term to the scalar potential, such that the two minima become non-degenerate. As in~\cite{Ferreira:2024eru}, we take
\begin{equation}
\label{eq:biasphi4}
V=V_{\mathbb{Z}_2}+ V_\text{bias}=\frac{\lambda}{4}\left(\phi^2-v^2\right)^2 + qv\phi^3,
\end{equation}
where $q\ll 1$ is a time-independent constant. We use a cubic, rather than linear, symmetry breaking potential to ensure that no bias is introduced in the initial population of the two minima.\footnote{Alternatively, one can choose a linear bias and set the initial homogeneous value of the field to coincide with the maximum of $V$, rather than setting $\phi_i=0$ as we do. We have checked that this alternative choice does not affect our results.}.  The potential~\eqref{eq:biasphi4} is shown by the blue curve in Fig.~\ref{fig:potentials}, for the actual value of $q$ used in our simulations. In the presence of $V_\text{bias}$, the energies of the two minima differ by $ \Delta V= q v^4 (4 \lambda^2 + 9 q^2)^{3/2} /(4 \lambda^3)$. It is useful to define a quantity $\eta_{\Delta V}$ as the (conformal) time at which the vacuum pressure equals the energy density of one Hubble sized wall, i.e. $\sigma H(\eta_{\Delta V})=\Delta V$. Roughly speaking, we expect the network to start deviating from the scaling regime around this time. 
This is confirmed by inspection of the area parameter, as shown in the left panel of Fig.~\ref{fig:areaparam}, where we have chosen $\eta_{\Delta V}=14$. The area parameter in the biased case coincides with its value in the unbiased case roughly until time $\eta_{\Delta V}$, then decreases as the network undergoes annihilation.

Two conditions restrict the range of bias sizes that we can simulate and the numerical parameters that are required to obtain reliable results. First, we need to ensure that the network follows the scaling regime for a while before the symmetry-breaking potential induces annihilation, otherwise our results would not be relevant to understand the signal from long-lived DWs. By activating a friction term between times $6.5 \leq \eta \leq 8$, we are able to accelerate the emergence of the scaling regime, which is then attained around $\eta \lesssim 14$, see App.~\ref{app:friction} for more details. Thus, the size of $q$ should be chosen so that $\eta_{\Delta V}\gtrsim 14$. On the other hand, the larger $\eta_{\Delta V}$ is, the longer the network needs to be simulated to ensure that annihilation and GW production are fully captured. The maximal time that can be reliably achieved is dictated by our computational capacity, mainly by the large amount of memory that is required to simulate on a lattice with a large number of lattice points. Given our available computational resources, simulating lattices much larger than $N=3060$ (including both GWs and fourth-order accurate spatial derivatives) is currently unfeasible.\footnote{Note that the higher the accuracy of the discretized spatial derivatives, the larger the number of `ghost' cells required by the simulation for parallelization purposes, and hence the larger the memory required.}.

Anticipating slightly the results of our simulations, we find that, for $\eta_{\Delta V}= 14$, GW production continues until very late times, around $\eta\approx 45$. According to the conditions in~\eqref{eq:tmax}-\eqref{eq:cmax}, by requiring $c(\eta_\text{max})\gtrsim 1$, and fixing $L\gtrsim 60$ in order to have sufficient IR resolution, these times are barely achievable with the box sizes that we are able to simulate. We are thus constrained to choose $\eta_{\Delta V}=14$ (we have nonetheless obtained results for the scalar spectrum with several other choices of bias size, see~App.~\ref{app:BiasSize}, that support the conclusion that the shape of the GW power spectrum should not be strongly affected by the bias size).

Nonetheless, we find that most of the walls disappear by the time $\eta\approx 36$, as can be appreciated in Fig.~\ref{fig:snapshots}, where we present snapshots of the gradient energy distribution at different times across a two-dimensional slice of the three-dimensional lattice. Correspondingly, we have indeed checked that the condition that the DWs width must be resolved until the end of the simulation is indeed overconservative: namely, comparing the results of a simulation with $L=80$ (where DWs are resolved until $\eta_{\rm max}=32$) to those with $L=60$ (where DWs are resolved until $\eta_{\rm max}=47$) we notice only marginal differences in our results. Therefore, we obtain most of our results with the parameters $N=3060, L=80$, which importantly provide a better IR coverage for the GW spectrum until late times. We do nonetheless test our results against individual realizations with smaller $L$, see App.~\ref{app:GWspec-LNdep} for more details.

The evolution of the energy density components of the scalar field is shown in the right panel of Fig.~\ref{fig:areaparam} for one particular realization of initial conditions. While the kinetic energy density deviates from its value in the unbiased case at around $\eta_{\Delta V}$ (gray vertical line), the gradient energy density decreases only at a later time. Gradients in the field dissipate once the network annihilates. At the latest times in the simulation shown in the right panel of Fig.~\ref{fig:areaparam}, the gradient energy density is comparable to the potential energy density. We confirm that the latter is diluted as non-relativistic matter at late times, $\sim a^{-3}$, as expected since it captures the oscillations of the scalar field around the true minimum. The gradient energy density instead dilutes faster, and we find it to provide the smallest contribution to the total energy density in longer simulations. We also find that after $\eta \approx 33$, the kinetic component approximately equals the sum of the gradient and potential components, as expected for a virialized field. On the other hand, the vacuum pressure due to the symmetry breaking term in the potential causes a late acceleration of the network, which induces a growth of the kinetic energy density until a time $\simeq 1.5\eta_{\Delta V}$, as can be appreciated in the right panel of Fig.~\ref{fig:areaparam}. Afterwards, the kinetic component decays; we find it to be diluted with the expansion slightly faster than non-relativistic matter but slower than radiation. While our simulations do not reach the regime where the kinetic energy density is comparable to the potential contribution, we expect this to be the case at later times, once the scalar field has cooled sufficiently to be entirely non-relativistic. Note that in all phenomenological applications with an observable GW signal one also needs to assume that the non-relativistic relics quickly decay into (dark or Standard Model) radiation, in order not to overclose the Universe (see e.g.~\cite{ZambujalFerreira:2021cte} for a discussion).

The snapshots in Fig.~\ref{fig:snapshots} allow to observe the formation, evolution, and annihilation of the DW network. As expected, the network achieves the scaling regime at $\eta \approx 14$, characterized by the existence of approximately one domain wall per Hubble patch. The emission of scalar waves by the DWs can also be appreciated, although these are still energetically subdominant at $\eta \approx 14$ (notice that the gradient energy density is plotted in logarithmic scale). After $\eta \geq 14$, the DW network starts annihilating, and the fraction of the lattice volume in the false vacuum starts diminishing, becoming negligible at $\eta \gtrsim 36$. Conversely, the energy density is redistributed in an inhomogeneous remnant fluid of scalar waves, that are initially mildly relativistic and are responsible for the slow decrease of the kinetic energy density in Fig.~\ref{fig:areaparam}.

\begin{figure}[t] 
    \begin{center}
        \includegraphics[width=0.47\textwidth]{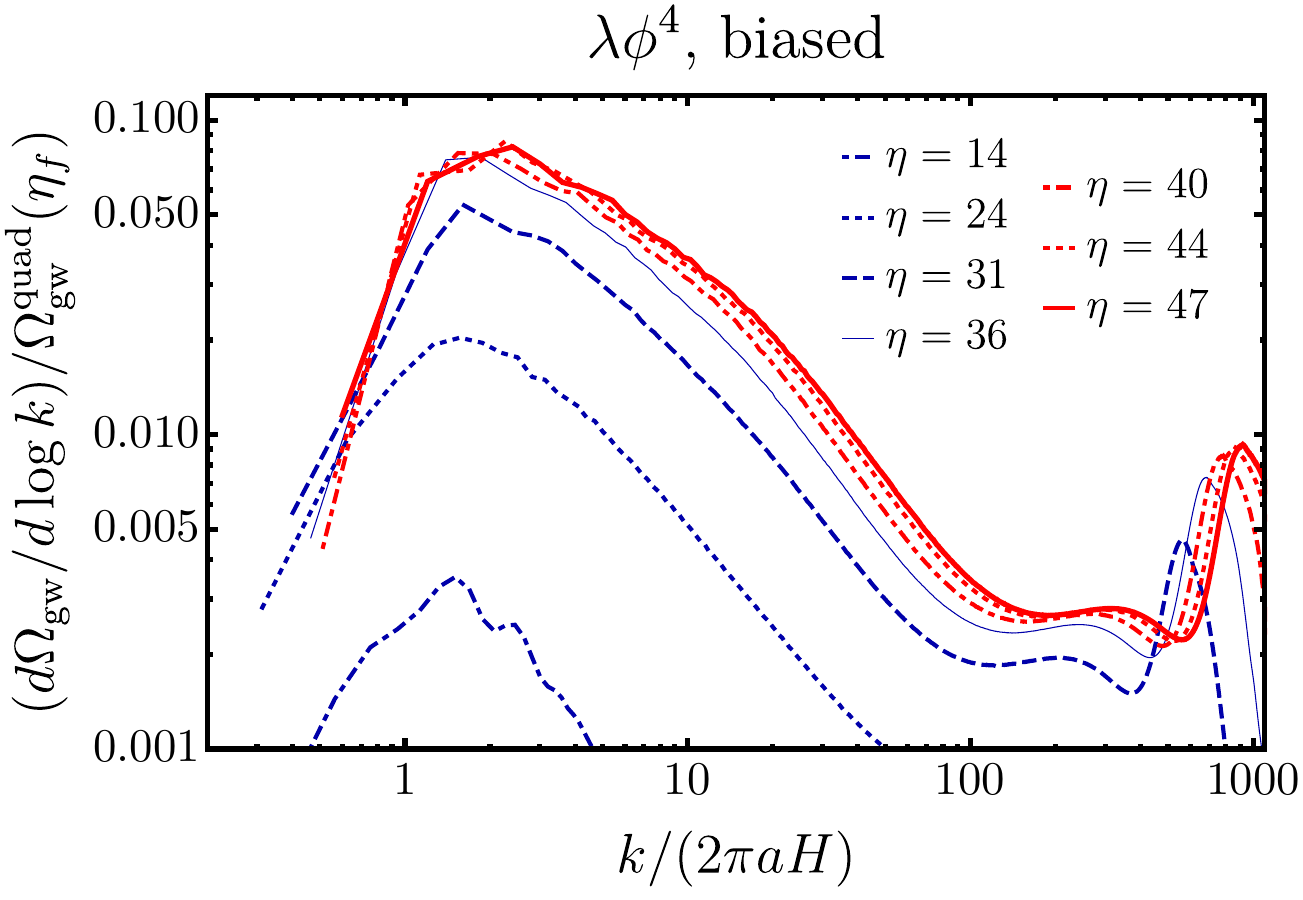}
    \end{center}
    \caption{Evolution of the GW spectrum from the collapsing network, obtained from averaging the results of seven realizations with different random initial conditions, with $N=3060$ and $L=80$. The thick solid red curve corresponds to the final time of the simulation $\eta_f = 47$, time at which the spectrum is fitted.}
\label{fig:spectra}
\end{figure}

Let us now move to the main aim of our work, namely the determination of the spectrum of GWs radiated by the evolving network. Due to the temporary growth of kinetic energy after $\eta_{\Delta V}$, the network efficiently radiates gravitational waves until a time $\eta_\text{gw} > \eta_{\Delta V}$. Our goal is to determine $\eta_\text{gw}$ and the GW spectrum at this time. The latter in particular exhibits some dependence on the initial conditions of the realization, with e.g.~changes of about 20\% in the amplitude of the main peak. We thus obtain the GW spectrum from seven realizations, and average over those to present a statistically reliable result at several times throughout the evolution. The resulting averaged GW spectra are shown in Fig.~\ref{fig:spectra}, 
with the spectrum at the final time shown by the solid red curve. For convenience, we have normalized the amplitude to the theoretical prediction from the quadrupole formula in the scaling regime with $\mathcal{A}=1$; i.e. to $\Omega^{\text {quad}}_{\text gw}\lvert_{\mathcal{A}=1}=3/(32 \pi) [2\sigma H/(3 H^2 M_p^2)]^2|_{\eta_f}$.\footnote{Notice that the quadrupole estimate of~\eqref{eq:quadrupole} is obtained by setting all time and length scales in the GW production process equal to $H^{-1}$. Nonetheless, our simulations with a bias show that the characteristic time/length scale of GW production is actually smaller than $H^{-1}$ by roughly a factor of two, in contrast with the result for a scaling network. Additionally, at late times GW production is dominated by scalar waves rather than by domain walls. Therefore, our parametrization is to be taken only as a convenient way to express our results, rather than as an estimate of the efficiency of GW production compared to the quadrupole formula.}

Several qualitative comments can be made before presenting a detailed fit of the signal. First, we notice that the position of the peak of the spectrum, $f_p$, at the final time deviates from the standard scaling results. This can be appreciated by comparing the spectrum at $\eta=14$ (when the symmetry breaking term is not active yet and the network is thus in scaling) and at $\eta = 47$. We find roughly $f_p (\eta=47) \approx 1.5 f_p (\eta\simeq 14)$. Secondly, the GW spectrum continues to grow until $\eta \approx 45$. Afterwards, the overall spectrum of the energy density fraction remains approximately constant, signaling that GW production has effectively stopped. Thirdly, the spectrum features a plateau-like region starting at $k/(2\pi a H)\gtrsim 100$, which is followed by a sharp peak close to the UV cutoff of the simulation. The latter feature is to be attributed to numerical effects, since its position and amplitude change for both different choices of lattice spacing and order of accuracy of the spatial derivatives, see e.g.~Fig.~\ref{fig:GradientOrderComp} in App.~\ref{app:spatialderivs}. It is less clear whether the plateau-like region has at least partially a physical origin (we will return to this point below, when considering other potentials). Nonetheless, we have checked that this region is always located close to the UV peak: therefore, even if it were physical, in a realistic cosmological context it would lie close to $k\sim m\gg H$ at the time of annihilation. In other words, the signal in this region would be suppressed by several orders of magnitude compared to the IR peak at  $k\sim 2\pi H$, and is thus most likely irrelevant for observational purposes.

\begin{figure}[t] 
    \begin{center}
        \includegraphics[width=0.47\textwidth]{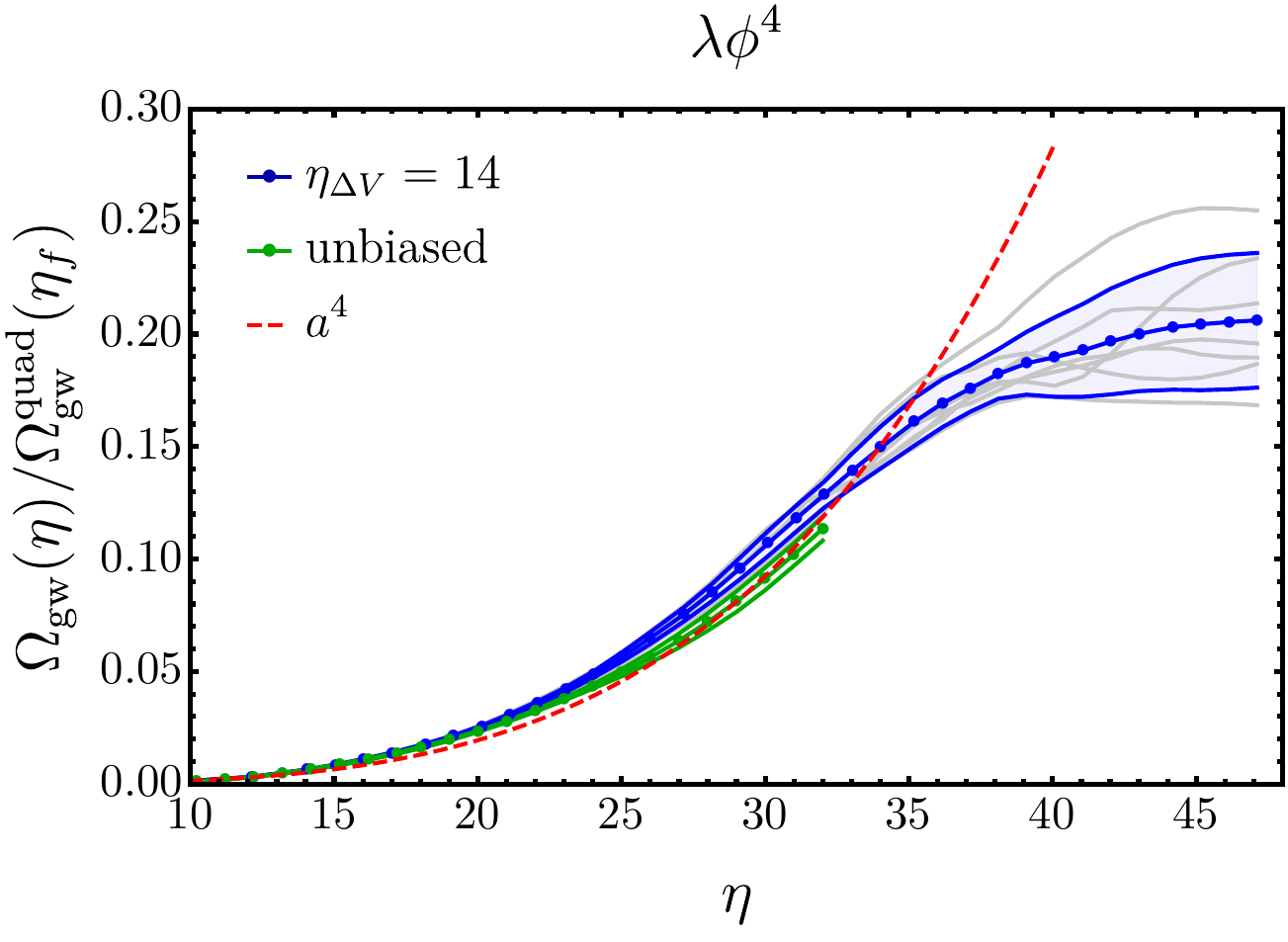}
    \end{center} \vspace{-0.2cm}
    \caption{Evolution of the integrated energy density fraction in GWs for biased networks, normalized to the theoretical prediction from the quadrupole formula in the scaling regime at time $\eta_f = 47$. Seven realizations are shown (gray curves), together with the average behavior (solid blue line with dots). The blue shaded region indicates the one standard deviation uncertainty. The green line and dashed region show the same quantities for the unbiased potential. The dashed red curve show the expected $\sim a^4$ behavior in the unbiased case.}
\label{fig:gwendensity}
\end{figure}

The integrated energy density fraction corresponding to the spectra in Fig.~\ref{fig:spectra} is shown by the solid blue curve in Fig.~\ref{fig:gwendensity}, with the shaded blue region indicating the one standard deviation uncertainty coming from averaging different realizations. As expected, we observe saturation of the signal around $\eta\approx 45$ (although the energy density fraction only grows by $\approx 7\% $ from $\eta\approx 40$ to $\eta\approx 45$). This can be contrasted with the unbiased case (green), with the same normalization as in the biased case, where the energy density fraction is observed to approximately grow as $a^{4}$ until the end of the simulation. While this is indeed the expectation from the scaling regime, we notice an intermediate regime in the unbiased case, roughly between $\eta\simeq 14$ and $\eta\simeq 30$, where the energy density fraction grows more slowly than $a^4$, which we interpret as a transient behavior, see App.~\ref{sec:unbiased} for more details. In the biased case, the energy density in GWs grows as $a^4$ only until $\eta\approx 27$.

After these preliminary comments, we are ready to present a fit to the average GW spectrum at the end of our simulations, i.e. $\eta_f=47$. We limit our analysis to wavenumbers $k/(2\pi a H)\leq 50$ for reliability purposes, as the spectrum for $x \geq 50$ shows a slight dependence on the choice of $L$, see App.~\ref{app:GWspec-LNdep} for more details. We parametrize the GW energy density spectrum at the final time of emission as follows:
\begin{equation}
\Omega_\text{gw}(k, \eta_f) = \Omega_\text{gw}(k_p, \eta_f)\times \mathcal{S}(x), \label{eq:GWspecParam}
\end{equation}
that is, we extract the amplitude of the signal at the peak wavenumber $k_p\gtrsim 2\pi a H$ and take the function $\mathcal{S}(x)$, where $x\equiv k/(2\pi a H)$, to describe only the spectral shape of the signal, by normalizing it to unity at $x_p=k_p/(2\pi a H)$. 

\begin{figure*}[t]
    \begin{center}
        \includegraphics[width=0.65\textwidth]{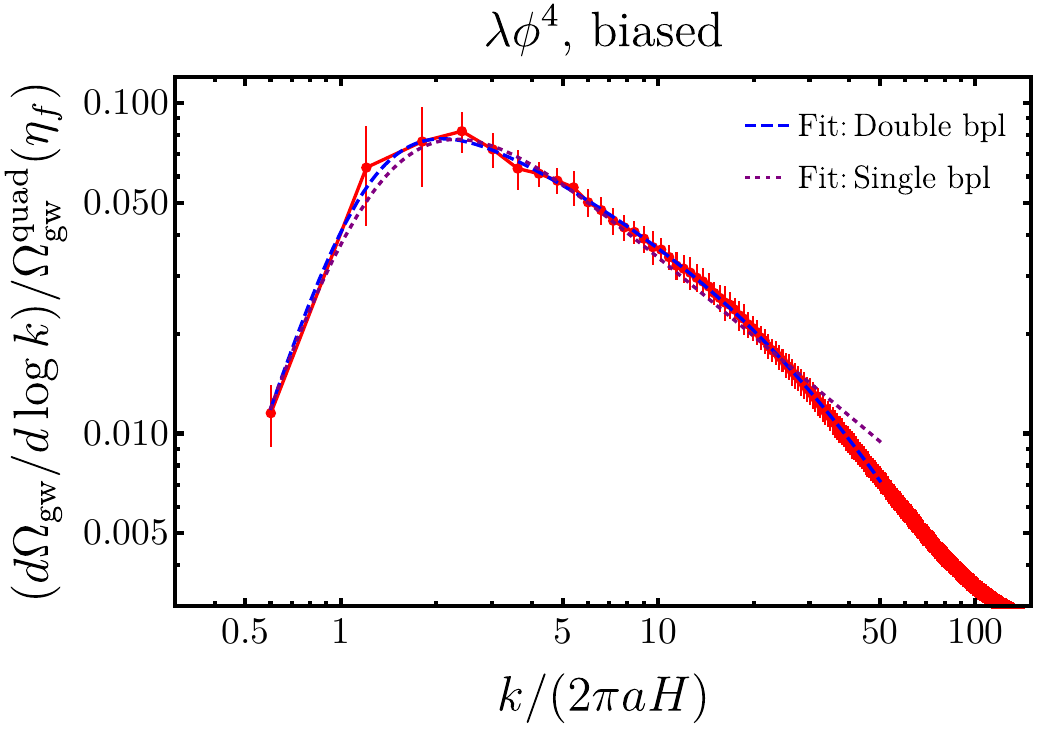}
    \end{center}
    \caption{Red dots show the amplitude of the averaged GW spectrum at different spectral bins for the biased $\phi^4$ potential, extracted at time $\eta_f = 47$, as a function of the wavenumber $x \equiv k / (2 \pi a H)$. The bars indicate the error coming from averaging seven different realizations. The blue/purple dashed lines show the best fit to the doubly/single broken power law templates, given by Eqs.~\eqref{eq:double-brokenpwl} and \eqref{eq:template}, respectively.}
\label{fig:lphi4-fitbias}
\end{figure*}

As for the spectral function $\mathcal{S}(x)$, we consider two possibilities. First, we fit with a standard peaked template (see e.g.~\cite{Caprini:2019egz}),
\begin{equation}
\label{eq:template}
\mathcal{S}(x)= \frac{(\alpha+\beta)^\delta}{\left( \beta \left( \frac{x}{x_p}\right)^{-\frac{\alpha}{\delta}} + \alpha \left( \frac{x}{x_p}\right)^{\frac{\beta}{\delta}}\right)^\delta} \ ,
\end{equation}
where $\alpha$ and $\beta$ capture the spectral slopes at $x\ll x_{p}$ and $x\gg x_{p}$ respectively, and $\delta$ is the width around the maximum. We set $\alpha=3$ because of causality \cite{Caprini:2009fx,Cai:2019cdl}. Overall, from the fit to the data we can extract the value of three parameters describing the shape of the spectrum, i.e.~$x_p$, $\beta$ and $\delta$, as well as the one additional parameter to describe its amplitude, the efficiency parameter $\epsilon$. We obtain the following constraints:

\begin{table}[H]
    \centering
    \begin{tabular}{|c|c|c|c|}
        \hline
        \multicolumn{4}{|c|}{{\bf Biased $\phi^4$ potential: single bpl template~\eqref{eq:template}}} \\ \hline
        $x_p$ & $\beta$ & $\delta$ & $\epsilon$ \\ \hline        
        $2.32 \pm 0.27$ & $0.80 \pm 0.08$  & $1.52 \pm 0.46$ & $0.078 \pm 0.007$ \\ \hline
    \end{tabular}
    \caption{Posteriors for the parameters of the single broken power law template \eqref{eq:template} for the biased $\phi^4$ potential.}
\end{table}

Secondly, we consider a peaked spectrum with a UV tail featuring a steepening of the slope at some wavenumber $k_b\gtrsim k_p$:
\begin{align} 
\mathcal{S} (x) = \frac{\alpha + \beta +(x_p/x_b)^{\beta+\gamma}}{\beta\left( \frac{x}{x_p} \right)^{-\alpha} + \alpha \left( \frac{x}{x_p} \right)^{\beta} + \left( \frac{x_b}{x_p} \right)^{-\beta} \left( \frac{x}{x_b} \right)^{\gamma} } \ ,\label{eq:double-brokenpwl} 
\end{align}
where $\alpha=3$, $\beta$ and $\gamma$ are the slopes at infrared, intermediate, and ultraviolet scales respectively, and $x_b = k_b /(2\pi a H)$. By fitting the data to this template, we now obtain constraints for four parameters describing the shape of the spectrum i.e.~$x_p$, $x_b$, $\beta$ and $\gamma$, as well as an additional one describing its amplitude, the efficiency parameter $\epsilon$. We obtain:
\begin{table}[H]
    \centering
    \begin{tabular}{|c|c|c|c|c|}
        \hline
        \multicolumn{5}{|c|}{{\bf Biased $\phi^4$ potential: doubly bpl template \eqref{eq:double-brokenpwl}}} \\ \hline
        $x_p$ & $x_b$ & $\beta$ & $\gamma$ & $\epsilon$  \\ \hline  
        $2.15 \pm 0.19$ & $6.2 \pm 1.9$  & $0.48 \pm 0.15$ & $1.79 \pm 0.36$ & $0.078 \pm 0.006$ \\ \hline 
    \end{tabular}
    \label{tab:fit-biasedlphi4-double}
    \caption{Posteriors for the parameters of the doubly broken power law template \eqref{eq:double-brokenpwl} for the biased $\phi^4$ potential, that provides an excellent fit to our numerical spectra.}
\end{table}

In Fig.~\ref{fig:lphi4-fitbias} we compare the results of both fits with the averaged GW spectrum extracted from the lattice. It can be appreciated that the peaked doubly broken power law template provides a significantly better fit to the numerical data than the simpler peaked template without a breaking at high frequencies. We then focus on the posteriors for the fitting parameters obtained with this spectral shape. 

Searches for GWs from domain walls have so far employed the single peaked template~\eqref{eq:template} with $\alpha=3$, mostly assuming $\beta=\delta=1$, following the results of~\cite{Hiramatsu:2013qaa}. Additionally, when translating the peak frequency of the signal to the corresponding temperature in the early Universe, it is commonly assumed that $x_p=1$. 
The GW spectrum obtained in our work is rather different in both aspects:
\begin{itemize}
\item[$\mathbf{1}$]{First, the GW signal from annihilating walls peaks at a wavenumber that is significantly larger than in the unbiased case. In particular, we find $x_p\simeq 2.2$, whereas in the scaling regime we find $x_p\simeq 1.2$, see App.~\ref{sec:unbiased}. This implies that the peak frequency of the GW signal today is roughly twice as large as the value that has been used so far in searches for GWs from domain walls in LVK and PTA datasets~\cite{Jiang:2022svq, Ferreira:2022zzo, NANOGrav:2023hvm}. We notice that this roughly agrees with the result presented in~\cite{Kitajima:2023cek}, which performed a single realization with a temperature-dependent linear bias to annihilate the network. In App.~\ref{sec:SpectraComp} we compare the results of~\cite{Kitajima:2023cek} with ours, while in Sec.~\ref{sec:variants} we consider the case of a temperature-dependent bias.}
\item[$\mathbf{2}$]{The spectrum is flatter in the intermediate region around the peak than in the scaling case. In the doubly broken power law model we find $\beta\approx 0.5$ up to a breaking wavenumber of roughly $x_b\approx 2.8~x_p$. Afterwards, we find a steeper slope of $\gamma\approx 1.8$.\footnote{This is valid in the range of $k$ that we have fitted to in our analysis. At very large $f$, one would expect the original scaling spectrum $f^{-1.2}$, see App.~\ref{sec:unbiased}, to eventually dominate.} The breaking point and the flatter slope was not appreciated in~\cite{Kitajima:2023cek}. Using instead the single broken power law spectrum, we find results that are in better agreement with previous analyses, i.e. $\beta\approx 0.9$, although the quality of the fit is significantly degraded.}
\item[$\mathbf{3}$]{If the single peaked template is employed, we find a width $\delta\simeq 1.5$, that is 50\% larger than the previously employed value.}
\item[$\mathbf{4}$]{The coefficient $\epsilon$ that measures the amplitude of the spectrum compared to a simple quadrupole estimate is measured to be $\epsilon\approx 0.08$, independently of the template that we used, with a very small uncertainty.} 

\end{itemize}

Before moving on to considering other DW models, let us discuss potential pitfalls of our approach and the associated potential impact on our results. For reasons of computational capabilities, we are not able to extensively test the validity of our results for different bias sizes (i.e. different values of $\eta_{\Delta V}$). One interesting and relevant question that requires further investigation is then whether the parameter $x_b$ actually depends on $\eta_{\Delta V}$. While we are not able to confidently address this question in this work, we notice that $x_b/x_p$ is close to the ratio between the time at which GW production starts deviating significantly from the scaling behavior (around $\eta\approx 28$, see Fig.~\ref{fig:gwendensity}) and the time at which production ends. We expect this ratio to remain roughly constant as we change $\eta_{\Delta V}$ (we have checked that this is the case with one other choice of bias size, and we have investigated the effects of different bias sizes on the scalar field spectrum, see~App.~\ref{app:BiasSize}, see also~\cite{Ferreira:2024eru}). Therefore, if the existence of the breaking point is related to the separation between $\eta_{\Delta V}$ and the final time of GW production, we expect $x_b$ to be roughly independent from $\eta_{\Delta V}$. The ratio $x_b/x_p$ is particularly relevant for observability at multiple GW observatories: if the main peak is indeed observed at e.g. PTAs, the UV tail would be observable at LISA if $x_b\gg x_p$, i.e. the intermediate slope extends for several orders of magnitude with $\beta\lesssim 1$. If instead the signal features a transition to a steeper slope at $x_b\simeq 2.8 ~x_p$ as we see in our simulations, then the UV tail would not be observable at LISA.

\section{Variants} \label{sec:variants}

Domain wall networks can arise in several different models, possibly with scalar fields that have a different potential from the quartic case that we have considered so far. Additionally, the symmetry breaking potential that induces annihilation can also differ from the time-independent cubic term that we have investigated in the previous section. It is thus interesting to ask whether variants of domain wall networks feature significant differences in their GW spectra from the case considered so far. This section aims to partially assess the model dependence of the GW spectrum, by considering two of the simplest and best-motivated variations of the domain wall network considered in Sec.~\ref{sec:biased}, both inspired by axion models. In particular, we first consider scalar field models where the potential (and the symmetry breaking term) is a cosine function. Separately, we consider the scenario where the symmetry breaking potential arises only at low temperatures, as is the case when it is induced by non-perturbative effects whose size grows rapidly close to confinement of some non-Abelian gauge sector.

\subsection{Cosine potential} \label{sec:Cos}

We start by considering a model with scalar potential 
\begin{equation}
\label{eq:cosine}
V=\lambda v^4 \left(1+\cos \frac{\pi \phi}{v}\right),
\end{equation}
that features an infinite number of degenerate minima, the first two of which are located at $\phi=\pm v$, exactly like in the $\phi^4$ case. The mass of the field around those minima is $m=v \sqrt{\lambda}\pi$, so that we can set again $m=v=1$ by fixing $\lambda=1/\pi^2$. We study the scenario in which the field is initially located at the maximum $\phi/v=0$, with random fluctuations of the same type as in Sec.~\ref{sec:biased} such that only the two nearest minima are initially populated. The resulting domain walls have tension $\sigma=8 v^3\sqrt{\lambda}/\pi$. When setting $m=v=1$, this gives a tension that is approximately $20\%$ larger than in the $\phi^4$ case, while the wall width is the same in the two models. Therefore, the resulting domain wall networks carries more energy density than in Sec.~\ref{sec:biased}.

\begin{figure*}[t]
        \includegraphics[width=0.47\textwidth]{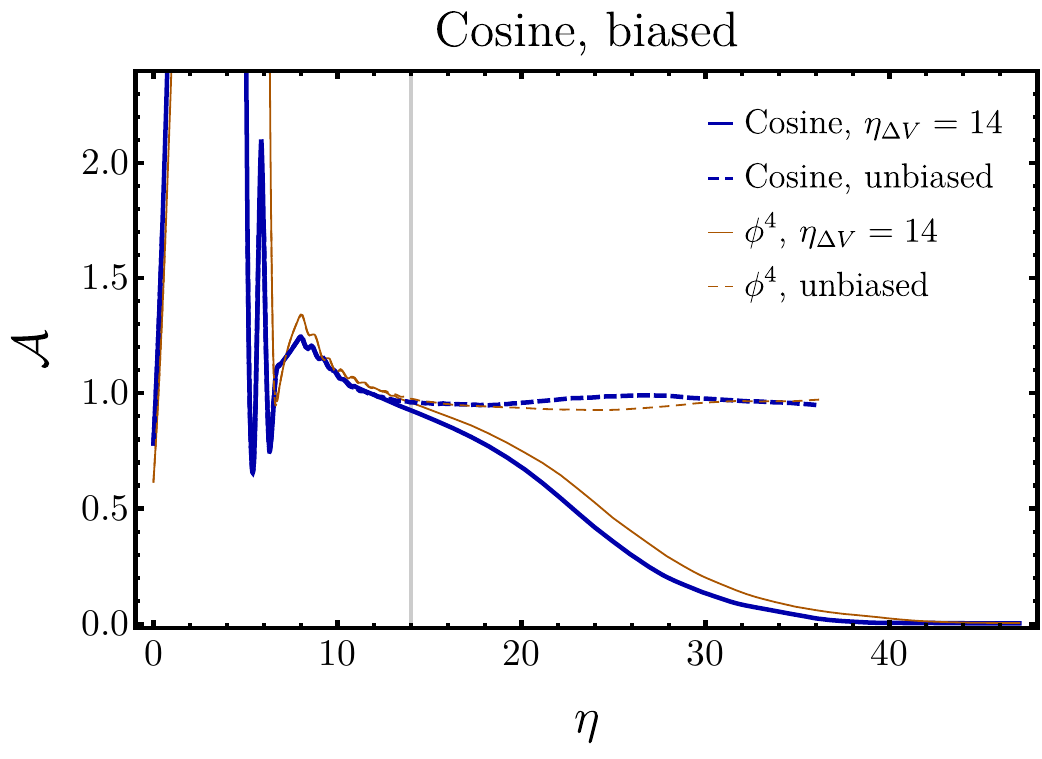}
        \hspace{1em}
        \includegraphics[width=0.46\textwidth]{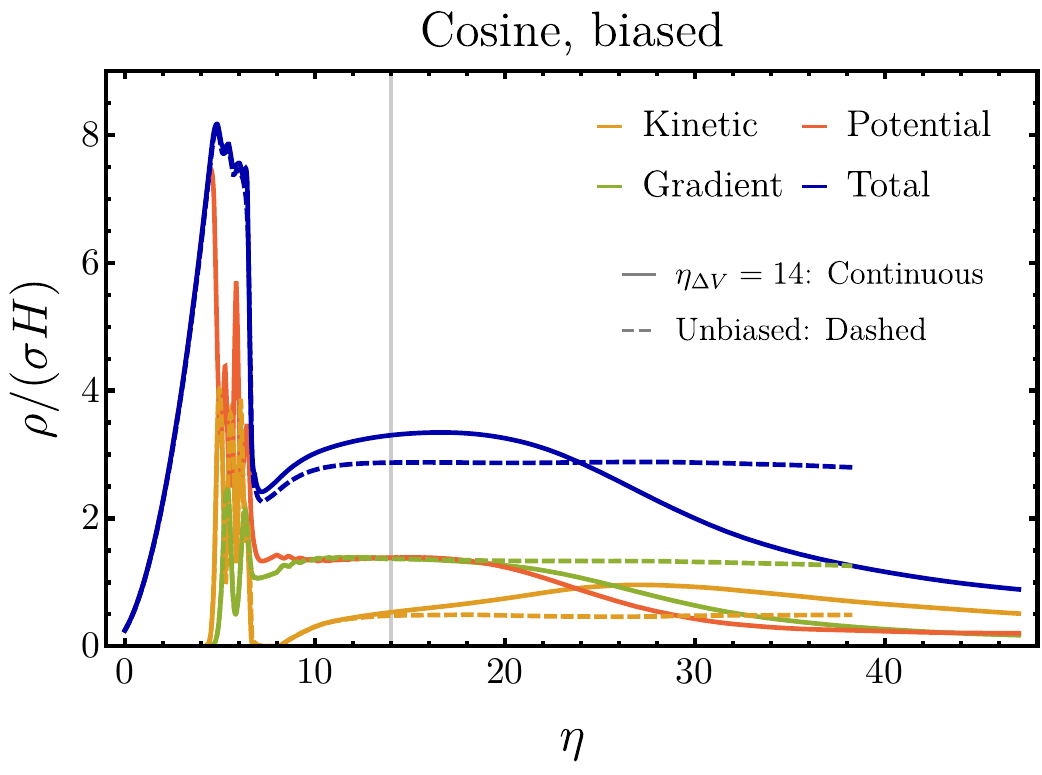} \vspace{-0.2cm}
        \caption{{\it Left:} Evolution of the area parameter for the biased and unbiased cosine potentials, obtained from simulations with $N=3060$ and $L=80$. The corresponding evolution for the biased and unbiased $\phi^4$ potentials is also shown for comparison. {\it Right:} Evolution of the energy density components of the scalar field for the biased and unbiased cosine potentials.} \label{fig:cosfigs} 
\end{figure*}

Since only two minima of the potential \eqref{eq:cosine} are initially populated, the area parameter of the resulting network in the absence of a symmetry breaking term closely resembles the analogous quantity in the $\phi^4$ potential, as can be appreciated in the left panel of Fig.~\ref{fig:cosfigs} (dashed curves). A similar remark applies to the components of the scalar field's energy density, shown in the top-right panel of Fig.~\ref{fig:cosfigs} (dashed curves).

In order to induce annihilation of the network, we introduce the following symmetry breaking term
\begin{equation}
\label{eq:biascos}
V_\text{bias}=p\lambda v^4\left[1+\cos\left(\frac{\pi\phi}{2v}-\frac{\pi}{2}\right)\right],
\end{equation}
whose periodicity is twice as large as that of potential~\eqref{eq:cosine}. The symmetry breaking potential is out of phase with respect to~\eqref{eq:cosine}, in such a way that the minimum at $\phi=-v$ is unaltered, while the minimum at $\phi=+v$ is lifted by $\Delta V \simeq 2p\lambda v^4$. The total potential is shown in Fig.~\ref{fig:potentials} by the orange curve. The conformal time $\eta_{\Delta V}$ is then determined by setting $\sigma H= \Delta V$ as usual. The potential \eqref{eq:biascos} induces a shift in the position of the maximum of the potential, differently from the $\phi^4$ case with the cubic bias. In order to avoid a bias in the initial population of the two minima, we thus set initial conditions for $\phi$ such that its homogeneous initial value coincides with the maximum of the total potential $V=V+ V_\text{bias}$, rather than setting $\phi=0$ as in Sec.~\ref{sec:biased}.

We simulate the network formation and annihilation using the same choice as in the previous case, i.e. $\eta_{\Delta V}=14$, with lattice parameters $N=3060$ and $L=80$. The behaviors of the area parameter and the components of the energy density, are shown by solid curves in Fig.~\ref{fig:cosfigs}. They closely resemble their analogous quantities in the $\phi^4$ case, with the most noticeable difference being that annihilation is slightly anticipated in the cosine case with respect to the $\phi^4$ case.

\begin{figure}[t]
    \centering
    \includegraphics[width=0.47\textwidth]{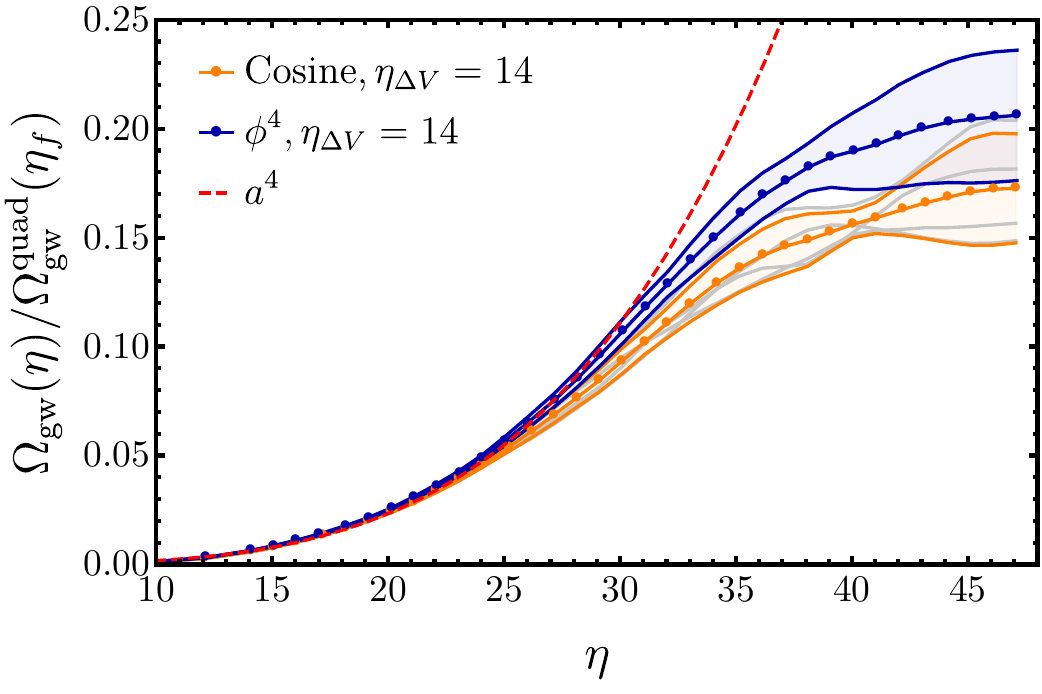}
    \caption{Evolution of the integrated energy density fraction in GWs for biased networks, showing both individual realizations for the cosine potential (gray curves) and the averaged result (dotted orange line). The shaded region corresponds to one standard deviation. The blue dotted line and shaded region show the averaged result for the $\phi^4$ potential for comparative purposes. The dashed red curve shows the expected evolution for unbiased potentials.}
    \label{fig:cosfigsB} 
\end{figure}

As in Sec.~\ref{sec:biased}, the detailed evolution of the GW spectrum for the biased cosine potential depends on the particular choice of random initial conditions. For that reason, we have carried out four simulations with different initial condition realizations. The integrated energy density fraction in GWs is shown in Fig.~\ref{fig:cosfigsB}. The dotted orange line shows the averaged evolution, while the shaded orange region corresponds to one standard deviation uncertainty. To the aim of comparing with the previous results, the same quantity in the $\phi^4$ case is also shown (blue shaded region and dotted line). As expected from the comparison of the area parameter and energy density components, the evolution of the GW energy density for the biased cosine potential is very similar to the $\phi^4$ case. The expected behavior in the scaling regime is shown by the red dashed line, which provides a good fit to the data until $\eta\approx 25$. The energy density fraction continues growing until eventually saturating at $\eta \approx 47$ (however, the growth from $\eta = 40$ until $\eta = 47$ is only $\sim 10\%$). The last stage of GW production can be attributed to scalar waves rather than domain walls like in the $\phi^4$ case. We also find $\Omega_{\rm gw} /\Omega_{\rm gw}^{\rm quad} \approx 0.17$, i.e.~$\sim 15\%$ smaller than in the $\phi^4$ case (notice that the domain wall tension is different in the two models, but both energy densities are normalized by their own quadrupole prediction in the scaling regime).

\begin{figure}[t]
    \centering
    \includegraphics[width=0.49\textwidth]{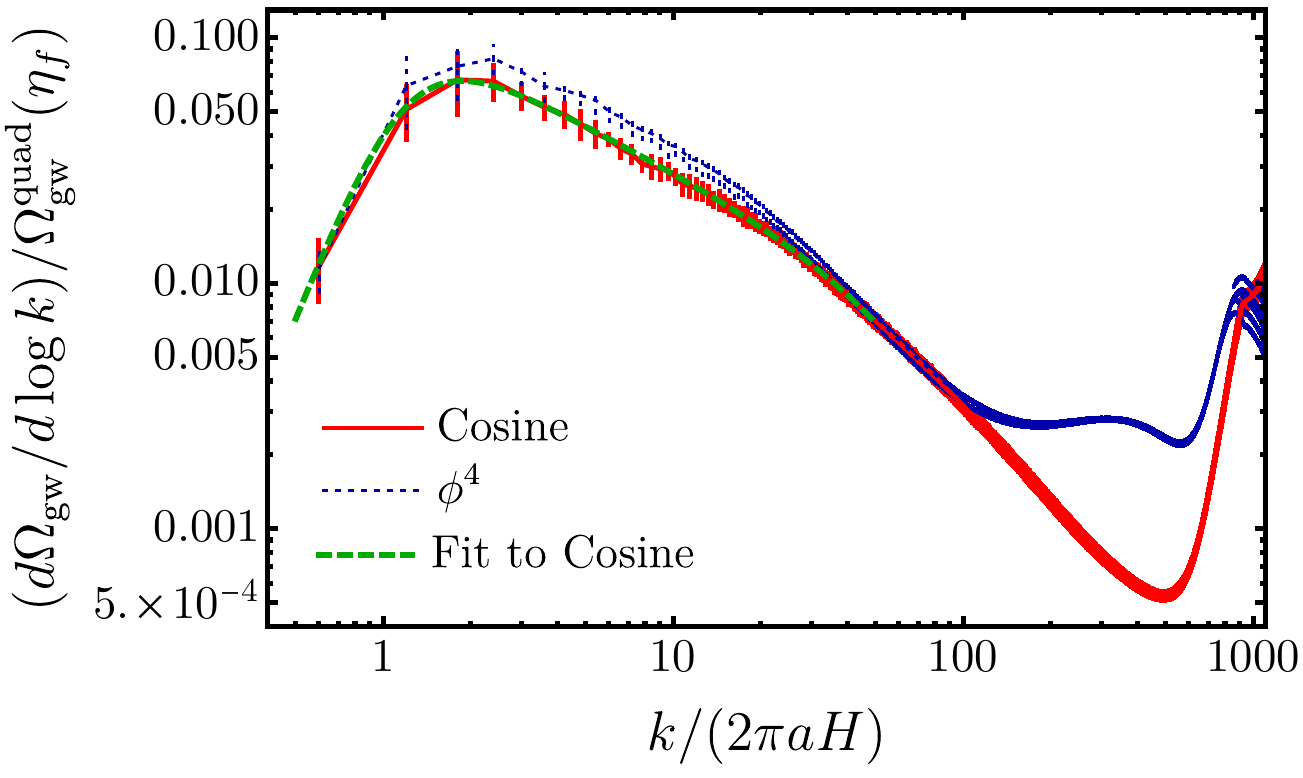}
    \caption{The red line shows the GW spectrum for the cosine potential at the final time $\eta_f = 47$, obtained by averaging the result from four simulations with different initial condition realizations. The green dashed line shows the best fit of this spectrum to template \eqref{eq:double-brokenpwl}. The blue line shows the averaged GW spectrum for the biased $\phi^4$ potential studied in Sec.~\ref{sec:biased}.}
    \label{fig:cosfigsC} 
    \end{figure}
 
The resulting GW spectrum at the final time $\eta_f = 47$ for the biased cosine potential is reported in Fig.~\ref{fig:cosfigsC}, averaged over four different realization of initial conditions. As in the $\phi^4$ case, we have fitted the final spectrum to the peaked broken power law template \eqref{eq:double-brokenpwl},  including only wavenumbers $x < 50$ in the fit. As seen in Fig.~\ref{fig:cosfigsC}, the function can describe quite well the shape of the spectrum around the peak. The fit yields the following estimation for the shape and efficiency parameters:

\begin{table}[H]
    \centering
    \begin{tabular}{|c|c|c|c|c|}
        \hline
        \multicolumn{5}{|c|}{{\bf Biased cosine potential: template \eqref{eq:double-brokenpwl}}} \\ \hline
        $x_p$ & $x_b$ & $\beta$ & $\gamma$ & $\epsilon$  \\ \hline  
        $ 1.88 \pm 0.06$ & $10.7 \pm 2.3$  & $0.61 \pm 0.04$ & $2.34 \pm 0.38$ & $0.067 \pm 0.002$ \\ \hline 
    \end{tabular}
    \label{tab:fit-biasedcos-double}
    \caption{Posteriors for the parameters of the doubly broken power law template \eqref{eq:double-brokenpwl} in the biased cosine case.}
\end{table}

In Fig.~\ref{fig:cosfigsC} we have also added, for comparative purposes, the final spectrum in the $\phi^4$ case reported in Sec.~\ref{sec:biased}. The amplitude of the final GW spectrum for the cosine potential is slightly smaller than in the $\phi^4$ case. Apart from that, the shape of both spectra are very similar, with only two noticeable differences. First, the breaking wavenumber in the cosine case is located at higher frequencies than in the $\phi^4$ case. Second, the behavior at $x \equiv k/(2\pi a H)>100$ is clearly different: in this region, the spectrum for the $\phi^4$ potential exhibits an intermediate plateau that is absent in the case of the cosine potential. We interpret this as being due to the different structure of the potentials at $\lvert\phi/v\lvert\gtrsim 1$: in the $\phi^4$ case the potential grows steeply, whereas the cosine potential has maxima followed by other minima.\footnote{It is in particular possible that the next-to-nearest minima are also populated during the annihilation process, due to the large release of kinetic energy \cite{Heilemann:2025iwv}. In our simulations, we have observed small regions of space where the field temporarily occupies the minimum at $\phi = -2v$ during the decay of the DW network. However, these regions are rare, and eventually settle at the $\phi = -v$ minimum.} However, as we have already mentioned, the details of such a UV part of the spectrum are most likely not relevant for observations, since this region always lies close to the inverse width of the wall, which in realistic scenarios is many orders of magnitude away from the main peak at $k\gtrsim 2\pi a H$.

\subsection{Temperature-dependent bias}  \label{sec:Tdep}

The second variation over the model presented in Sec.~\ref{sec:biased} consists in introducing a temperature dependence in the symmetry breaking potential $V_\text{bias}$ of \eqref{eq:biasphi4}. As mentioned above, this scenario captures those models in which annihilation occurs because the (pseudo)scalar field couples to some additional sector, that undergoes a phase transition at low temperatures. The scenario of reference is that of symmetry breaking due to non-perturbative effects in confining gauge sectors (see~\cite{Borsanyi:2016ksw} for the QCD axion case), though weakly coupled models may also exhibit the same phenomenon.

\begin{figure*} 
    \begin{center}
        \includegraphics[width=0.47\textwidth]{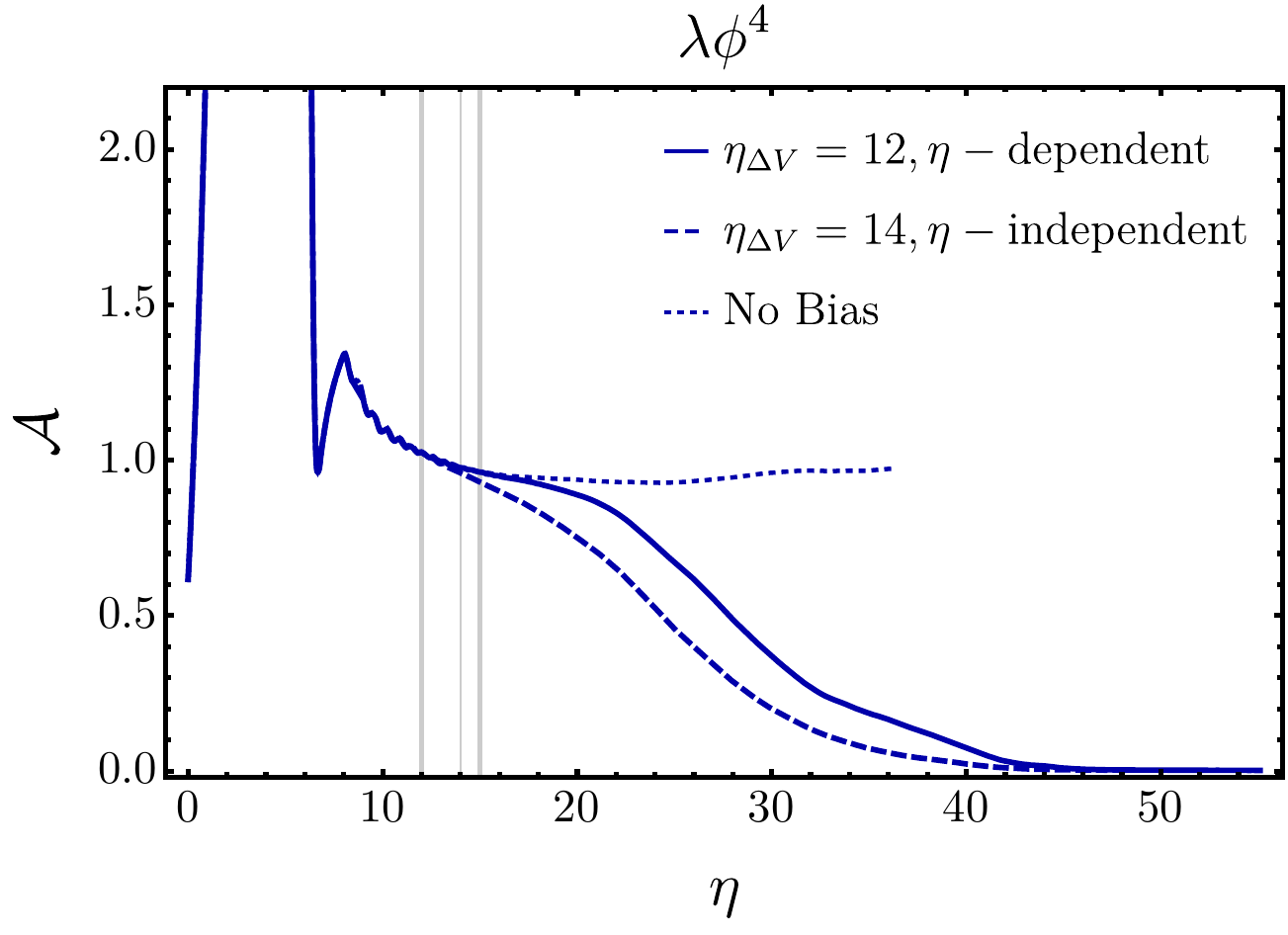}
        \hspace{1em}
        \includegraphics[width=0.46\textwidth]{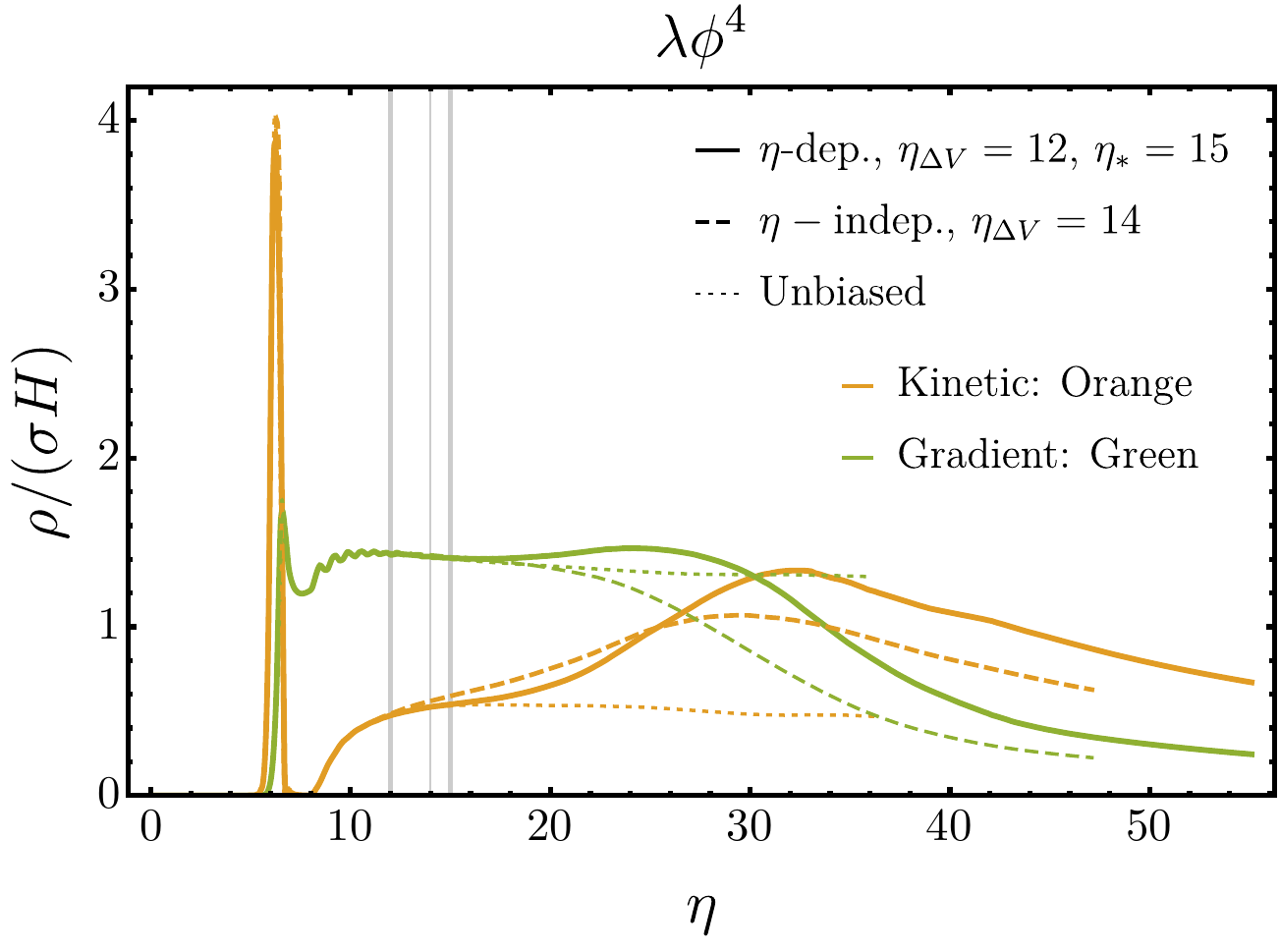}
        \end{center}
         \caption{{\it Left:} Evolution of the area parameter in the presence of a time dependent symmetry breaking term in the potential (solid curve), for $\eta_{\Delta V} = 12$, $\eta_{\star} = 15$, and obtained from a simulation with $N=3060$ and $L=70$. The evolution of the area parameter for a time-independent symmetry breaking term with $\eta_{\Delta V}=14$ (dashed curve) and for an unbiased potential (dotted curve) are also shown for comparison. {\it Right:} Evolution of the kinetic and gradient energy density components of the scalar field in the same three cases.} \label{fig:Tdepfigs}
   \end{figure*} 

The temperature dependence of $V_\text{bias}$ is in general model-dependent. To set ideas, here we consider the following example
\begin{equation}
V_\text{bias}(T)=\frac{qv\phi^3}{1+[a(\eta)/a(\eta_\star)]^{-b}},
\end{equation}
where $b>0$. For $\eta\ll \eta_\star$, the potential is then suppressed, while it saturates to the function considered in Sec.~\ref{sec:biased} for $\eta\gtrsim \eta_\star$. To fix ideas, we set $b=8$ in the rest of this section, which roughly reproduces the temperature dependence of the axion potential from QCD instantons~\cite{Borsanyi:2016ksw}, since $a=(1+\eta)\propto 1/T$ (although in that case the potential is periodic, rather than being a cubic monomial). In our numerical implementation of the model we translate the temperature dependence to a (conformal) time dependence, by means of $a(T)\sim T^{-1}\sim (1+\eta)^{-1}$. The energy difference between the two minima $\Delta V(\eta)$ then becomes time-dependent. As we shall see, in the presence of time dependence, annihilation is slightly delayed compared to the time independent case. Therefore, we are forced to consider a larger bias potential than in Sec.~\ref{sec:biased}, and in particular we fix $q$ such that $\sigma H=\Delta V(T=0)$ at $\eta_{\Delta V}=12$. For the same reason, we are forced to fix $\eta_\star\propto 1/T_\star$ to a value that is rather close to $\eta_{\Delta V}$, otherwise annihilation would complete only at times that are beyond the reach of our simulations. We thus set $\eta_\star=15$.  We set initial conditions as in Sec.~\ref{sec:biased}, and simulate the dynamics in lattices of $N=3060$ and $L=70$.

The behaviors of the area parameter (left panel of Fig.~\ref{fig:Tdepfigs}) and the kinetic and gradient energy densities (right panel) for one specific realization of initial conditions are shown by the solid curves in Fig.~\ref{fig:Tdepfigs}. For comparison, the same quantities for the time-independent model of Sec.~\ref{sec:biased} (dashed curves) and in the absence of a symmetry breaking term are also reported (dotted curves). Significant differences with the time-independent scenario can clearly be appreciated. In particular, the annihilation of the network is significantly delayed when using a time-dependent bias potential with $\eta_\star\gtrsim \eta_{\Delta V}$, as expected.

\begin{figure} 
    \begin{center}
        \includegraphics[width=0.47\textwidth]{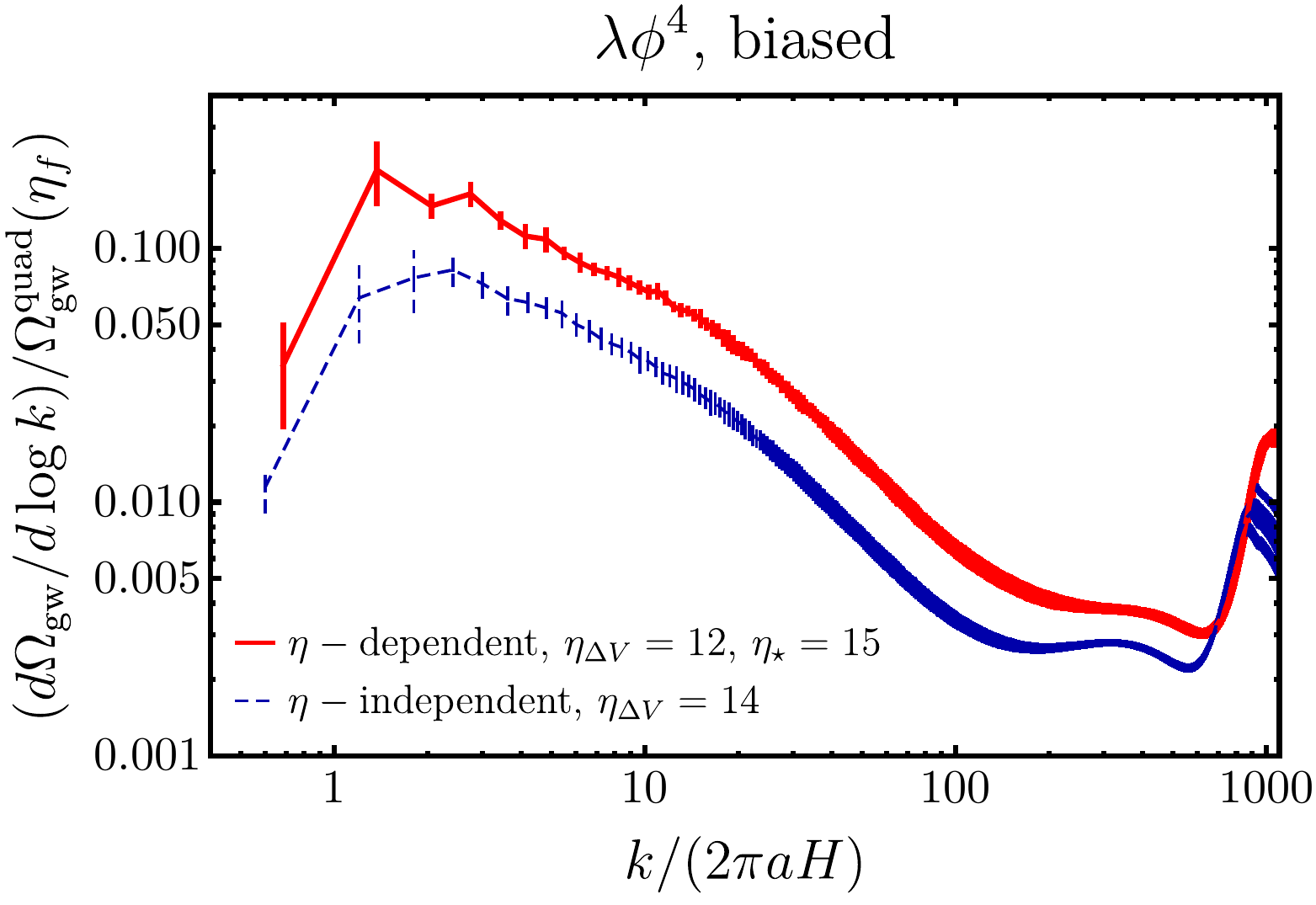}
        \end{center}
         \caption{Comparison of the GW spectra in the biased $\phi^4$ case at the final time $\eta_f = 47$ in the presence of either a time-dependent (solid red) or time-independent (dashed blue) symmetry breaking term.} \label{fig:TdepfigsB}
   \end{figure} 

We have carried out three simulations of the time-dependent bias case with different initial conditions realizations. Accidentally, for our particular choices of $\eta_{\Delta V}$ and $\eta_\star$, we find that GW production ceases to be efficient around $\eta \approx 45$, approximately at the same time as in the time-independent bias case studied in Sec.~\ref{sec:biased}. However, we stress that this result is obtained with a larger bias than in Sec.~\ref{sec:biased}; with the same bias size, GW production would continue until later times.

In Fig.~\ref{fig:TdepfigsB}, the GW spectrum at the final time $\eta_f=47$ is compared with the result of Sec.~\ref{sec:biased} at the same time. We find that the amplitude of the GW spectrum is approximately a factor two larger than with a time-independent bias, with the efficiency parameter being now $\epsilon = 0.159 \pm 0.006$. Apart from that, the shapes of the spectrum in both cases are very similar.

\section{Conclusions and Observational Impact}
\label{sec:conclusions}

Domain wall networks in the early Universe have been recently receiving significant attention, in particular because of their Gravitational Wave (GW) radiation.

Previous work has shown that the GW relic abundance from domain wall networks is set by their annihilation phase, which is required by observational viability, rather than by the scaling regime as previously thought. However, a characterization of the GW spectrum in this regime had so far been missing. This work fills this gap and provides the GW community (especially the observational collaborations) with the state-of-the-art GW spectrum from cosmologically viable DW networks.

Our results have been achieved by simulating DW network dynamics in scalar field simulations, and extracting the corresponding GW radiation, in lattices of up to $N=3060$ points per dimension, as well as used other numerical techniques -an early period of friction and higher-order accurate spatial derivatives- to maximize the accuracy of the results extracted from the lattice. 

Our main result is that the relic GW spectrum from cosmic domain walls is very well modeled as a peaked function, whose high-frequency tail is a broken power law with two distinct behaviors. In particular, the energy density spectrum today is
\begin{equation}
\Omega_\text{gw}(f) = \Omega_\text{gw}(f_p)\times \mathcal{S}(f/f_p), \label{eq:GWtoday}
\end{equation}
where the spectral shape function is 
\begin{align} 
\label{eq:shape2}
\mathcal{S} (x) &= \frac{3 + \beta +(f_p/f_b)^{\beta+\gamma}}{\beta\left( \frac{f}{f_p} \right)^{-3} + 3 \left( \frac{f}{f_p} \right)^{\beta} + \left( \frac{f_b}{f_p} \right)^{-\beta} \left( \frac{f}{f_b} \right)^{\gamma} } \ ,
\end{align}
with $\beta\simeq 0.5$, $\gamma\simeq 1.8$ (see Table~\ref{tab:fit-biasedlphi4-double} for detailed posteriors). In previous literature the peak frequency has been fixed to the frequency $f_\text{gw}$, corresponding to the Hubble scale at a given temperature $T_\text{gw}$ in the early Universe, at which GW production ceases to be efficient, see below for further details (and e.g. the appendix of~\cite{Franciolini:2023wjm} for a derivation of the relation between frequencies and temperature). Our work finds instead that $f_p$ is roughly twice as large as $f_\text{gw}$, more precisely:
\begin{align}
\label{eq:pfreqtoday}
\nonumber f_p &\simeq 24~x_p~\text{nHz}\left(\frac{g_*(T_\text{gw})}{100}\right)^{1/6}\left(\frac{T_\text{gw}}{150~\text{MeV}}\right) \ , \\
x_p&=2.15 \pm 0.19 \ ,
\end{align}
where $g_*(T_\text{gw})$ is the number of relativistic degrees of freedom at the temperature $T_\text{gw}$ (and we have simplified by setting $g_{*,s}(T_\text{gw})=g_*(T_\text{gw})$).  

Our result~\eqref{eq:shape2} shows that the spectrum decreases very slowly in an intermediate frequency range between the peak and the UV breaking frequency $f_b$, as $\sim f^{-\beta}\simeq f^{-0.5}$, in contrast to the previously employed value $\beta= 1$. However, such an intermediate region of the spectrum extends only up to $f\gtrsim f_b\simeq 2.8 f_p$. At higher frequencies, the spectrum decreases more steeply, roughly as $f^{-1.8}$.

Concerning the amplitude of the signal, our results can be incorporated in the standard estimate for the GW relic abundance from domain walls (see e.g.~\cite{Ferreira:2022zzo}) as:

\begin{equation}
\Omega_\text{gw}(f_p)h^2\simeq 10^{-10}\epsilon\left(\frac{g_*(T_\text{gw})}{10.75}\right)^{-1/3}\left(\frac{\alpha_\text{gw}}{0.01}\right)^{2}.
\end{equation}
The new result of our work in this expression is $\epsilon\simeq 0.07-0.08$ (depending on the potential, see below), with the definition $\alpha_\text{gw}\equiv 2\sigma H_\text{gw}/(3H_\text{gw}^2M_p^2)$ and $H_\text{gw}\equiv H(T_\text{gw})$.

Additionally, we were able to improve on the determination of $T_\text{gw}$, i.e. the temperature at which efficient GW production stops (and thus the temperature that sets the position of the peak frequency of the relic GW spectrum~\eqref{eq:pfreqtoday} as well as the amplitude of $\alpha_\star$), in terms of the DW tension $\sigma$ and the vacuum energy difference between the two minima $\Delta V$. Previous work~\cite{Ferreira:2024eru} has already found that $T_\text{gw}< T_{\Delta V}$, where $T_{\Delta V}$ is the temperature defined by the commonly employed condition $\sigma H=\Delta V$ for DW annihilation. In our simulations with $\eta_{\Delta V}=14$ we find $\eta_\text{gw}=44.4\pm 1.0$, i.e. $\eta_\text{gw}\simeq 3.2\eta_{\Delta V}$. Therefore, we find an even larger delay in GW production with respect to what was already noticed in~\cite{Ferreira:2024eru}, attributing the differences to the improvements in our simulations. A careful extension of this result to a general bias size requires further studies, although we expect the amount of the delay to remain constant. Under the assumptions that the delay remains constant when changing the size of the bias term (see~\cite{Ferreira:2024eru} and our App.~\ref{app:BiasSize}), we deduce\footnote{In~\cite{Ferreira:2024eru}, the parameter $\eta_{\text{ann}}$ was introduced, corresponding to the time at which the fraction of the simulation box in the false vacuum (i.e.~in $\phi>0$) starts decreasing exponentially. It was found that $\eta_\text{ann}\gtrsim \eta_{\Delta V}$. One would expect $\eta_\text{gw}/\eta_\text{ann}$ to be independent of $\Delta V$, rather than $\eta_\text{gw}/\eta_{\Delta V}$. Nonetheless, if $\eta_\text{ann}$ and $\eta_{\Delta V}\propto \Delta V^{1/2}$ have the same dependence on $\Delta V$, then the extrapolation~\eqref{eq:tgw} is  justified. By fitting the results presented in Table 1 of~\cite{Ferreira:2024eru}, we find that $\eta_{\rm ann} \propto (\Delta V)^{-(0.40 \pm 0.05)}$.  This result is compatible up to two standard deviations with the standard expectation, although it is also compatible with the recent claim by~\cite{Babichev:2025stm} $\eta_{\rm ann} \propto (\Delta V)^{1/3}$ (that appeared after our work, and is obtained with lower resolution than~\cite{Ferreira:2024eru}). While a dedicated analysis is required to reach a conclusion, we notice that a different scaling would only affect the relation between the amplitude of the GW signal and the parameters $\sigma$ and $\Delta V$. The shape of the spectrum would not be altered.}
\begin{align}
\label{eq:tgw}
&T_\text{gw}\simeq 0.3 T_{\Delta V}\\&\simeq 133~\text{MeV}\left(\frac{g_*(T_\text{gw})}{17.25}\right)^{-\frac{1}{4}}\left(\frac{\Delta V^{1/4}}{100~\text{MeV}}\right)^{2}\left(\frac{10^5~\text{GeV}}{\sigma^{1/3}}\right)^{\frac{3}{2}} \, . \nonumber 
\end{align}

The new properties of the GW spectrum that we have uncovered are important when searching for DWs in GW datasets. At present, evidence of a stochastic GW background exists in PTA datasets and thus our spectrum is particularly relevant for model comparison within those data. Understanding the impact of our result for the goodness of the fit to PTA data, and for the extraction of domain wall parameters, requires a detailed Bayesian search that updates existing results~\cite{Ferreira:2022zzo, NANOGrav:2023hvm}, which we leave to future work. With respect to current data, we can anticipate one dominant effect: the shift in the peak frequency with respect to the previously employed values affects the preferred temperature of GW production $T_\text{gw}$, which we also find to be roughly a factor of $3$ smaller than what previously used in GW searches. Thus, the inferred posteriors for DW parameters $\sigma$ and $\Delta V$ will be affected. On the other hand, the shallower slope around the peak that we have found should have a milder impact on the fit to current data, since those prefer the IR tail of cosmological spectra.

We have found our conclusions to be mostly rather robust against the choice of the potential (assuming that only two minima are initially populated), as well as the time-dependence of the annihilation-inducing bias, for the scenarios that we have tested. Specifically, we have obtained results for a periodic potential, of the form that arises in axion models. While axion models have richer defect networks than the one studied here, we expect our results to apply to the axion scenario with $N_\text{dw}=2$, as also found in previous work~\cite{Hiramatsu:2012sc}. Extending our analysis to axionic string-wall network is an interesting task for future work. Concerning the symmetry breaking potential, we have obtained results for a time-dependent bias using a power-law parametrization with the exponent that applies to the QCD axion case. In both cases, the shape of the GW spectrum is very similar to the case of $\phi^4$ potential with time-independent bias. Nonetheless, we do find a larger amplitude of GWs in the time-dependent bias case, roughly by a factor of two.
Extending our results to other time-dependent behaviors, that arise in specific models, may be of interest.

\bigskip

{\it Note added:} shortly before our paper appeared on the arXiv, the preprint~\cite{Cyr:2025nzf} has also presented results on gravitational waves from annihilating domain walls, based on lattice simulations of $N=2048$ points per dimension.

\acknowledgments

The authors thankfully acknowledge the use of computational resources from the Nyx cluster at ICCUB, the LXPLUS cluster at CERN, and Finisterrae III at CESGA. The work of F.T.~is supported by a \textit{Beatriu de Pinós} fellowship 2022-BP-00063 from the Ministry of Research of Catalonia. The work of F.R.~is supported by the grant RYC2021-031105-I from the Ministerio de Ciencia e Innovación (Spain). The work of
A.N. and F.T. is supported by the grants 
PID2022-137268NB-C52 from the Spanish Ministry of Science and Innovation, Unit of Excellence Maria de Maeztu 2024-2027 of ICCUB (CEX2024-001451-M) and AGAUR 2021 SGR 00872. We thank Jorge Baeza-Ballesteros for his valuable assistance in producing snapshots of the lattice. We acknowledge fruitful interactions during completion of this work at the VIII Cosmological Olentzero Workshop in Bilbao, and thank the UPV for the warm hospitality.

\appendix

\section{Lattice simulations for the unbiased $\phi^4$ potential} \label{sec:unbiased}

In this Appendix we present results from lattice simulations for the $\phi^4$ potential \eqref{eq:quartic-pot} in the absence of a symmetry break term, which complement the ones for biased potentials presented in the bulk text. In this scenario, the domain walls remain in the scaling regime indefinitely, with approximately one domain wall per Hubble patch.

\begin{figure*}[!p]
    \centering

    \begin{minipage}[t]{\textwidth}
        \centering
        \includegraphics[width=0.48\textwidth]{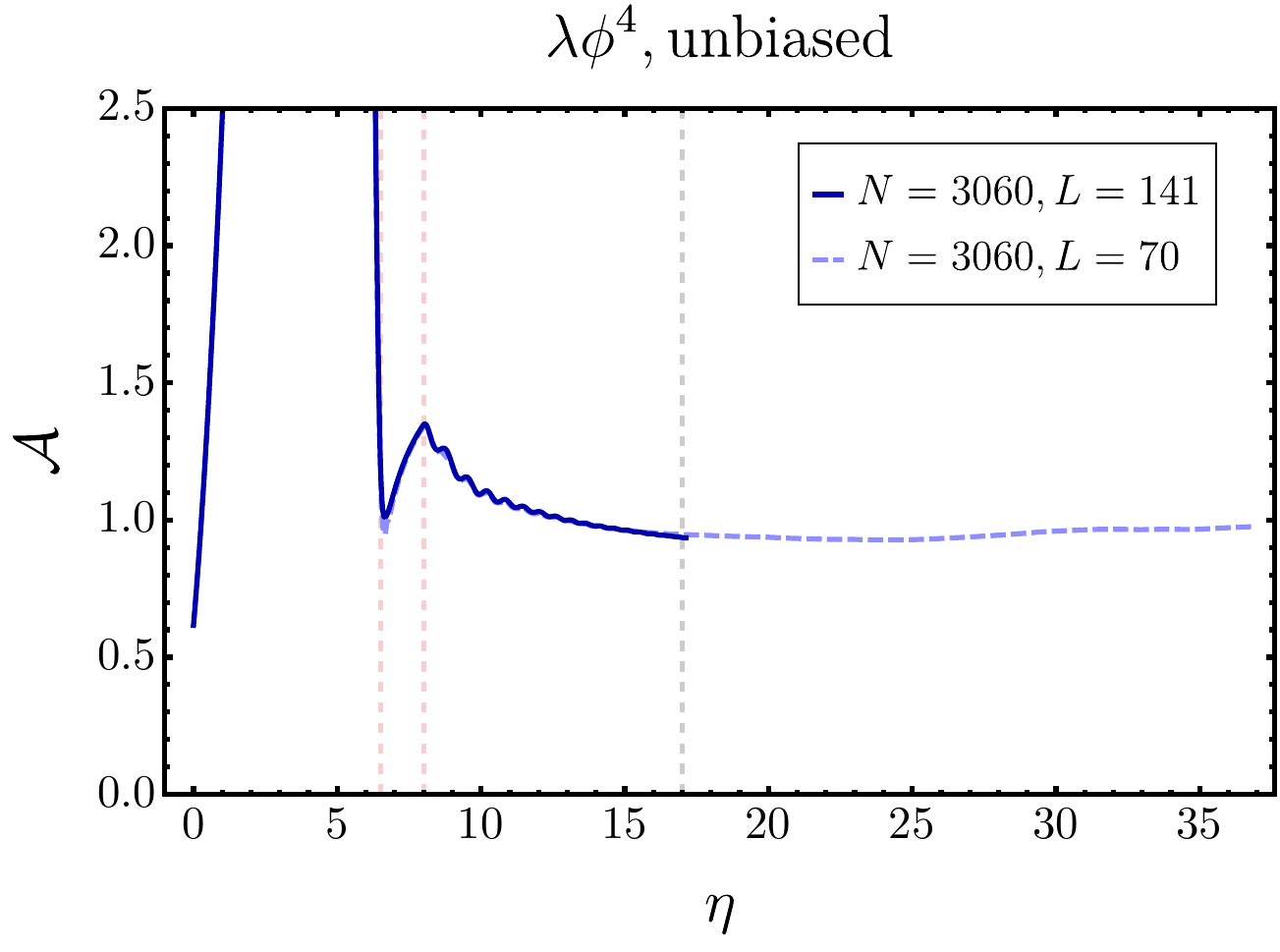} \hspace{1em}
    \includegraphics[width=0.48\textwidth]{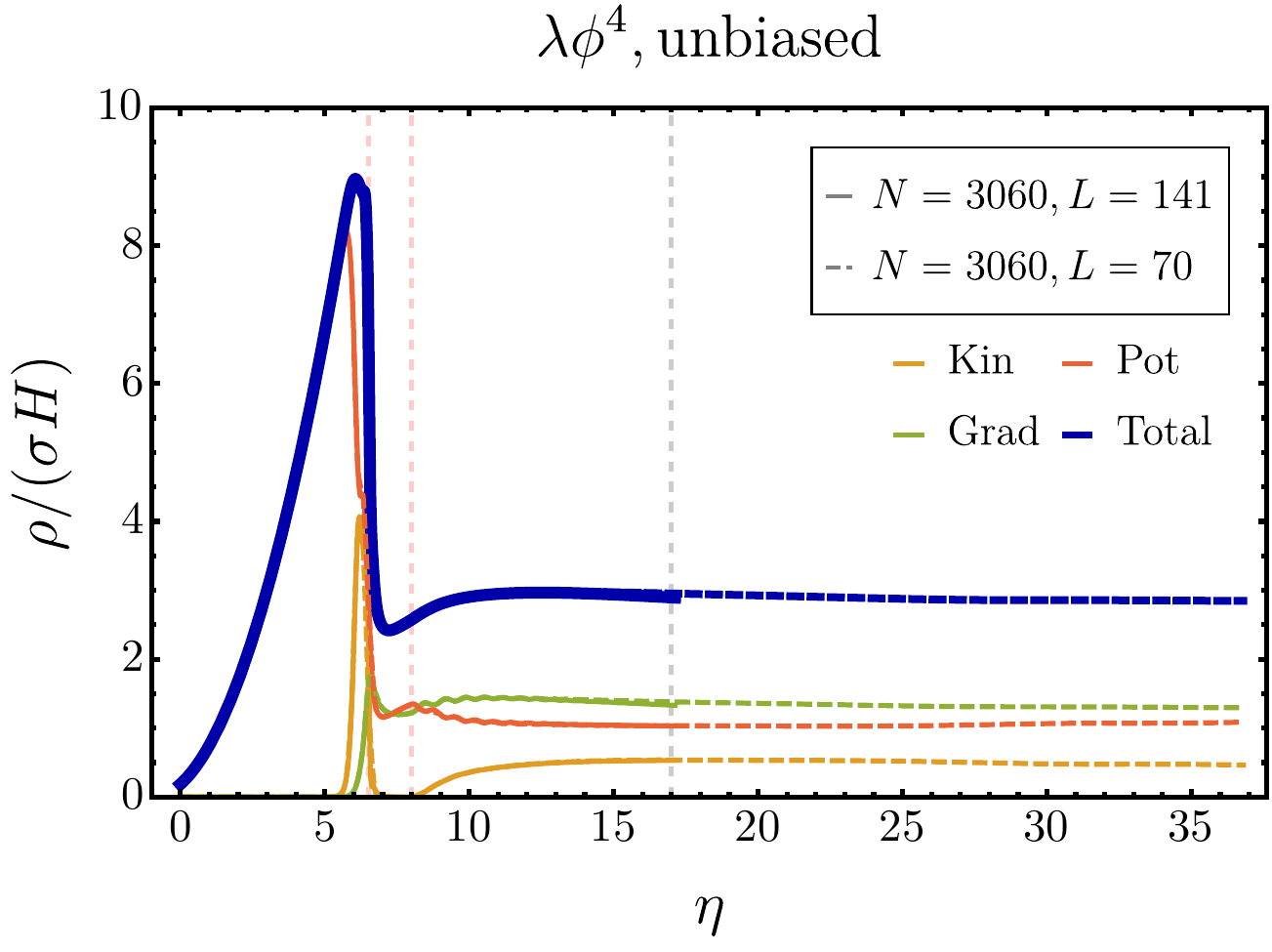}
        \caption{\textit{Left:} Evolution of the area parameter for the $\phi^4$ potential in the absence of a symmetry breaking term, extracted from two simulations with $N=3060$ and $L=141,\,70$. The red vertical dashed lines delimit the time interval in which the friction term is active, while the gray one indicates the time at which the $L=141$ simulation stops being reliable. \textit{Right:} Evolution of the total energy density and its kinetic, gradient, and potential contributions for the same two simulations.}
        \label{fig:DynamicsFrict}
    \end{minipage} 

    \vspace{1.5cm}

    \begin{minipage}[t]{0.48\textwidth}
        \centering
        \includegraphics[width=\textwidth]{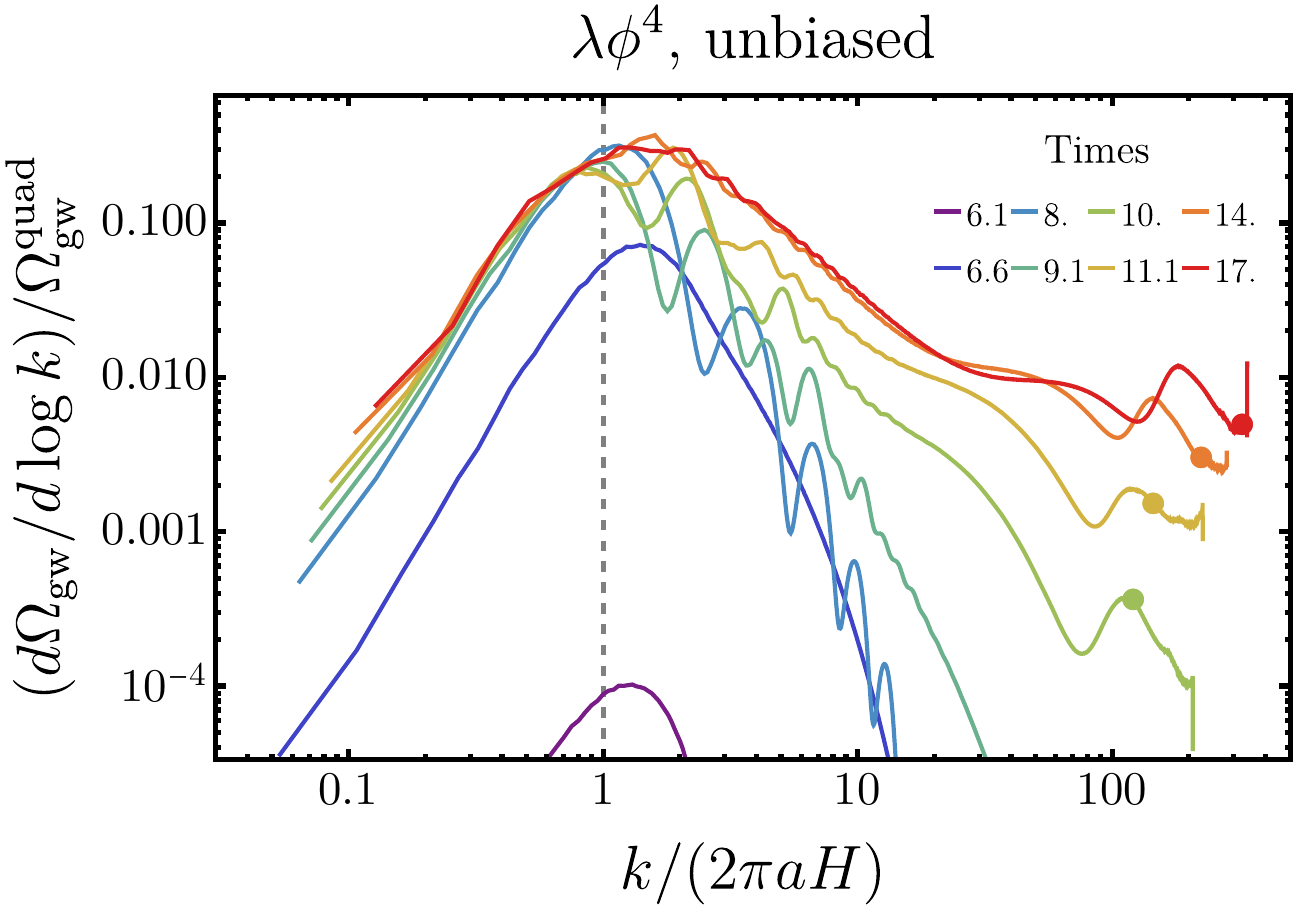}
        \caption{Evolution of the GW spectrum generated during the formation and early scaling regimes of the DW network, obtained with a simulation for the unbiased $\phi^4$ potential for lattice parameters $N=3060$ and $L=141$. The dots over the lines indicate the momentum scale associated to the DW width at each time.}
        \label{fig:GWScSpectra}
    \end{minipage}
    \hfill
    \begin{minipage}[t]{0.48\textwidth}
        \centering
        \includegraphics[width=\textwidth]{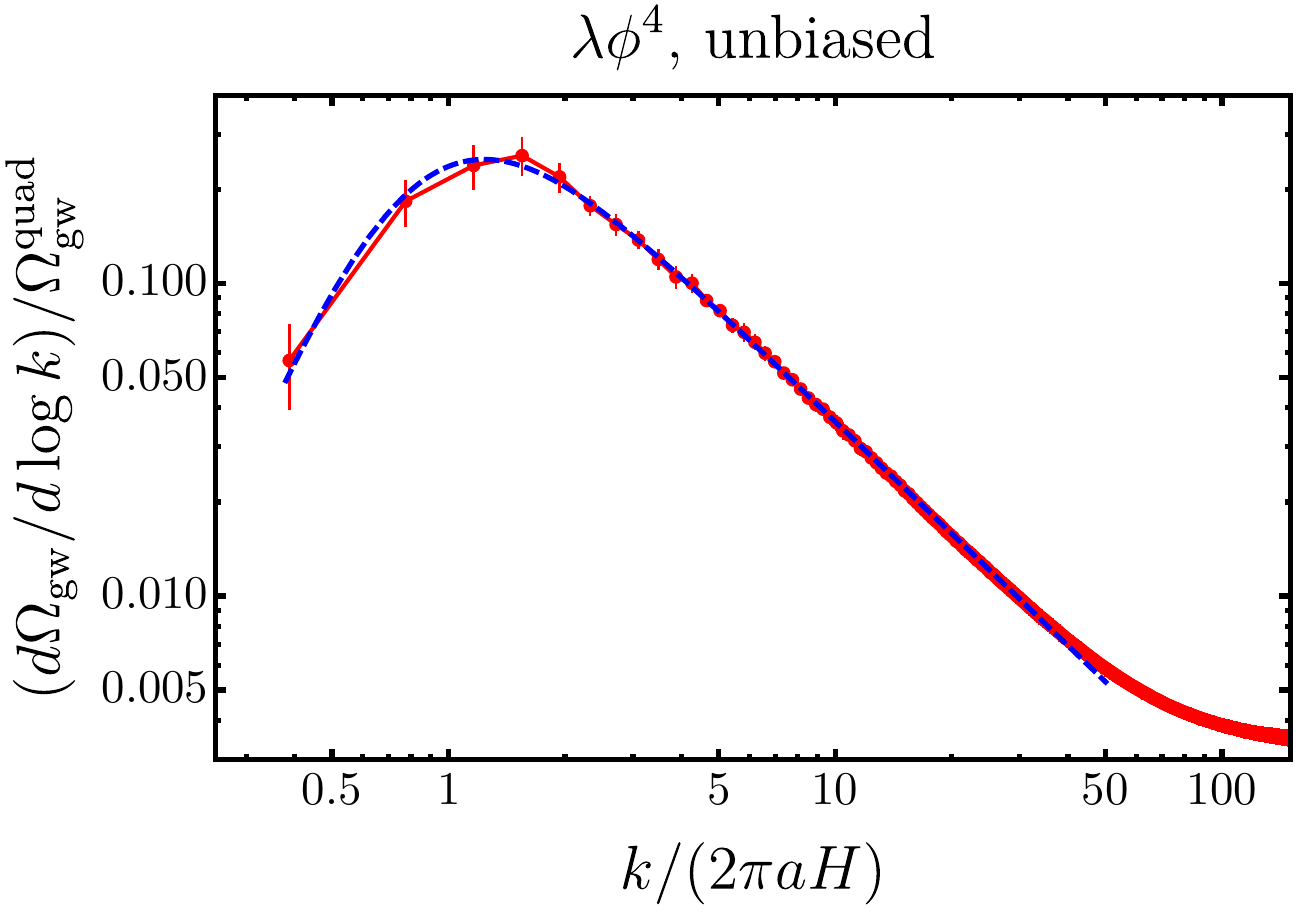}
        \caption{Red dots show the GW spectrum amplitude at each spectral bin at the time $\eta_f = 30$, as a function of the wavenumber $x \equiv k /(2\pi a H)$, obtained by averaging the results of five simulations with lattice parameters $N=3060$, $L=80$ and different realizations for the initial conditions. The error bars show one standard deviation uncertainty at each bin. The dashed blue line shows the best fit to the function \eqref{eq:template}.}
        \label{fig:lphi4-scaling-fit}
    \end{minipage}

\end{figure*}

In Fig.~\ref{fig:DynamicsFrict} we compare the evolution of the area parameter and energy density components for two simulations of the unbiased case, with lattice parameters $N=3060$ and $L=141,\,70$. In both cases we have activated the friction term at times $6.5 \leq \eta \leq 8$. The first simulation only allows to evolve the domain walls until the time $\eta_{\rm max} \approx 18$, but the large lattice volume allows to capture many Hubble patches, down to $\sim\text{420}$ at the end of the simulation. On the contrary, the second simulation allows to simulate the DWs until a larger time, $\eta_{\rm max} \approx 36$, but captures less Hubble patches, down to $\sim\text{6}$ when the simulation ends. Despite these differences, we observe that the evolution of both the area parameter and energy density components coincide almost exactly at the times when both simulations are valid, confirming the consistency of our results. We also observe that after friction is deactivated, the area parameter quickly settles towards a value slightly less than one, attained approximately at the time $\eta \approx 14$. This time scale characterizes the establishment of the scaling regime. Regarding the energy densities, we observe that the friction triggers a sharp decrease of the total energy density, coming from a loss of both potential energy (because the field settles at the minima of the potential) and kinetic energy (because the walls freeze). After friction gets deactivated, the energy components quickly converge towards the values $\rho (\eta_{\rm sc} ) \simeq 2.9 \, \sigma H (\eta_{\rm sc})$, with the kinetic, gradient, and potential contributions representing $18\%$, $47\%$ and $35\%$ of the total energy density respectively. These values remain approximately constant until the end of the simulation.

In Fig.~\ref{fig:GWScSpectra} we show the evolution of the scalar field spectra during the formation and early scaling regime of the DW network, obtained from the $N=3060$, $L=141$ simulation. As expected, the scalar field spectrum quickly develops a main peak approximately at the wavenumber $x \equiv k/(2\pi a H) \simeq 1$, which gets imprinted in the GW spectrum at the slightly higher frequency $ \simeq 1.3 - 1.4$. To the left of the peaks, the GW spectrum behaves approximately as $\sim k^{3}$ as expected. To the right of the peak, the GW spectrum at $\eta = 17$ decreases as a power-law at scales $1.4 < x < 20$, while it develops a plateau for $20 <x <100$. Note also that a peak appears in the GW spectrum at $x \approx 200$, which is just an artifact coming from lack of UV resolution in the lattice, similar to the one found for biased potentials. As observed in Fig.~\ref{fig:GWScSpectra}, the GW spectrum saturates approximately when the scaling regime is established at the time $\eta \approx 14$.

Let us now present a fit of the GW spectrum for the unbiased quartic potential. For such purpose, it is convenient to parametrize the region around the peak as in Eq.~\eqref{eq:GWspecParam}, where $k_p$ is the comoving momentum of the spectrum's peak, $\mathcal{S} (x)$ with $x \equiv k / (2\pi a H)$ is the shape function, normalized to unity at $x_p \equiv k_p/(2 \pi a H)$, and $\eta_f$ is now a time after the GW spectrum stops evolving. We approximate the shape function by the single peak template \eqref{eq:template}. We have simulated the system in lattices of $N=3060$ and $L=80$ five times, in order to account for the dependence of the GW spectrum on the random initial fluctuations. In our simulations, we have observed that the amplitude of the averaged GW spectrum at $k=k_p$ slightly decreases after $\eta \geq 17$, eventually saturating at $\eta \approx 30$, with $\Omega_{\rm gw} (k_p, \eta = 30) \approx 0.8\, \Omega_{\rm gw} (k_p, \eta = 17)$. The observed loss of amplitude is likely due to a transient production of GWs associated with the formation of domain walls, which gradually dilute as radiation after scaling regime is achieved. Therefore, we carry out the fit of the GW spectrum at the later time $\eta_f = 30$, when it saturates. We depict the resulting averaged GW spectrum at that time in Fig.~\ref{fig:lphi4-scaling-fit}, with the errors bars indicating the one standard deviation uncertainty in each spectral bin. We then fit the averaged spectrum to the single peak template \eqref{eq:template}, considering only wavenumbers $x \leq 50$. We obtain the following constraints for the shape and efficiency parameters:
\begin{table}[H]
   \centering
   \begin{tabular}{|c|c|c|c|} \hline
    \multicolumn{4}{|c|}{{\bf Unbiased $\phi^4$ potential: single peak template \eqref{eq:template}}} \\
        \hline
        $x_p$ & $\beta$ & $\delta$ & $\epsilon$ \\ \hline  
        $1.24 \pm 0.06$ & $1.19 \pm 0.02$  & $1.69 \pm 0.28$ & $0.249 \pm 0.017$ \\ \hline
    \end{tabular}
    \caption{Posteriors for the parameters of the single peak template in the case of the unbiased $\phi^4$ potential.}
\end{table}

Our result for the spectrum in the scaling regime can be compared with those found in previous (lower-resolution) simulations~\cite{Hiramatsu:2013qaa, Dankovsky:2024zvs}, whereas the results of~\cite{Ferreira:2023jbu} have been obtained assuming a matter-dominated background. First, we notice that the width $\delta$ was not included as a parameter in~\cite{Hiramatsu:2013qaa, Ferreira:2023jbu, Dankovsky:2024zvs}. Additionally,~\cite{Hiramatsu:2013qaa, Dankovsky:2024zvs} only provide fits to the high frequency region of the spectrum, rather than a fit to the full spectrum.
Concerning the slope of the spectrum in this range ($f\geq f_p$), our results are close to the usually adopted value $\beta\approx 1$, derived in~\cite{Hiramatsu:2013qaa}, and also agree with~\cite{Dankovsky:2024zvs}, whereas~\cite{Ferreira:2023jbu} finds a significantly steeper decrease in matter domination (with white noise initial conditions as in our case). Concerning the position of the peak, we also find reasonable agreement with the expectation $x_p\simeq 1$. Finally, concerning the amplitude parameter $\epsilon$, our result is within the errors of~\cite{Hiramatsu:2013qaa} and agrees very well with~\cite{Dankovsky:2024zvs}.

\clearpage

\newpage
\section{Further numerical details} \label{app:technicalities}

In this Appendix we provide supplementary material about our lattice results, including more details about the implementation of initial random fluctuations, friction, and spatial derivatives of higher accuracy. We also assess the robustness of our results for the GW spectrum by analyzing its dependence on the lattice volume and bias term, as well as compare with previous results in the literature.

\subsection{Initial conditions}

We have evolved the fields in a radiation-dominated background $a(\eta) = a_i + H_i \eta$, with $a_i$ and $H_i$ the scale factor and Hubble parameter at the initial simulation time respectively. At that time we set the homogeneous mode of the inflaton amplitude to zero. Over it, we impose a white noise spectrum of Gaussian fluctuations in momentum space below a certain cutoff $k< k_*$. The variance in position space is given by  
\begin{align} \langle \delta \phi^2 \rangle = & \int d {\rm log} k \, \Delta_{ \phi} (k) \ , \hspace{0.4cm} \Delta_{\phi} (k) = c \frac{k^3}{4 \pi^2 m} \Theta (k- k_*) \ , \label{eq:InitPowSpec} \end{align} 
where we have set a tiny initial amplitude with an arbitrary small coefficient  $c=10^{-10}$, since the scaling solution should not depend on it, and $k_*$ a ultraviolet cutoff. The procedure by which this spectrum is generated is explained in Sec.~7.1 of \cite{Figueroa:2020rrl}. In our simulations we have fixed $k_* = 2 m$, but we have checked that the dynamics of the DW network are quite insensitive on the specific choice of $k_*$.

\begin{figure*}[t]
    \begin{center}
        \includegraphics[width=0.46\textwidth]{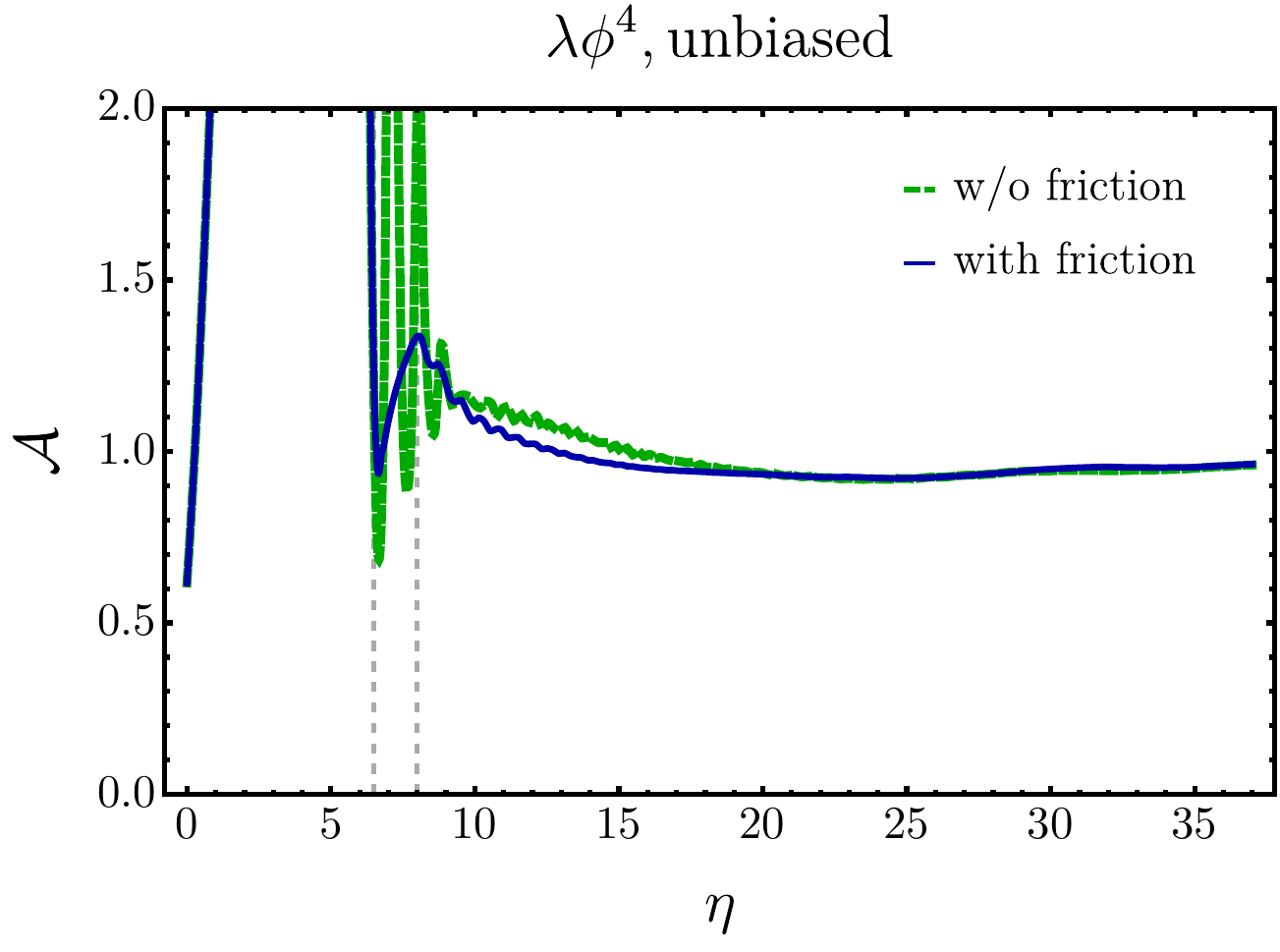} \,\,
        \includegraphics[width=0.46\textwidth]{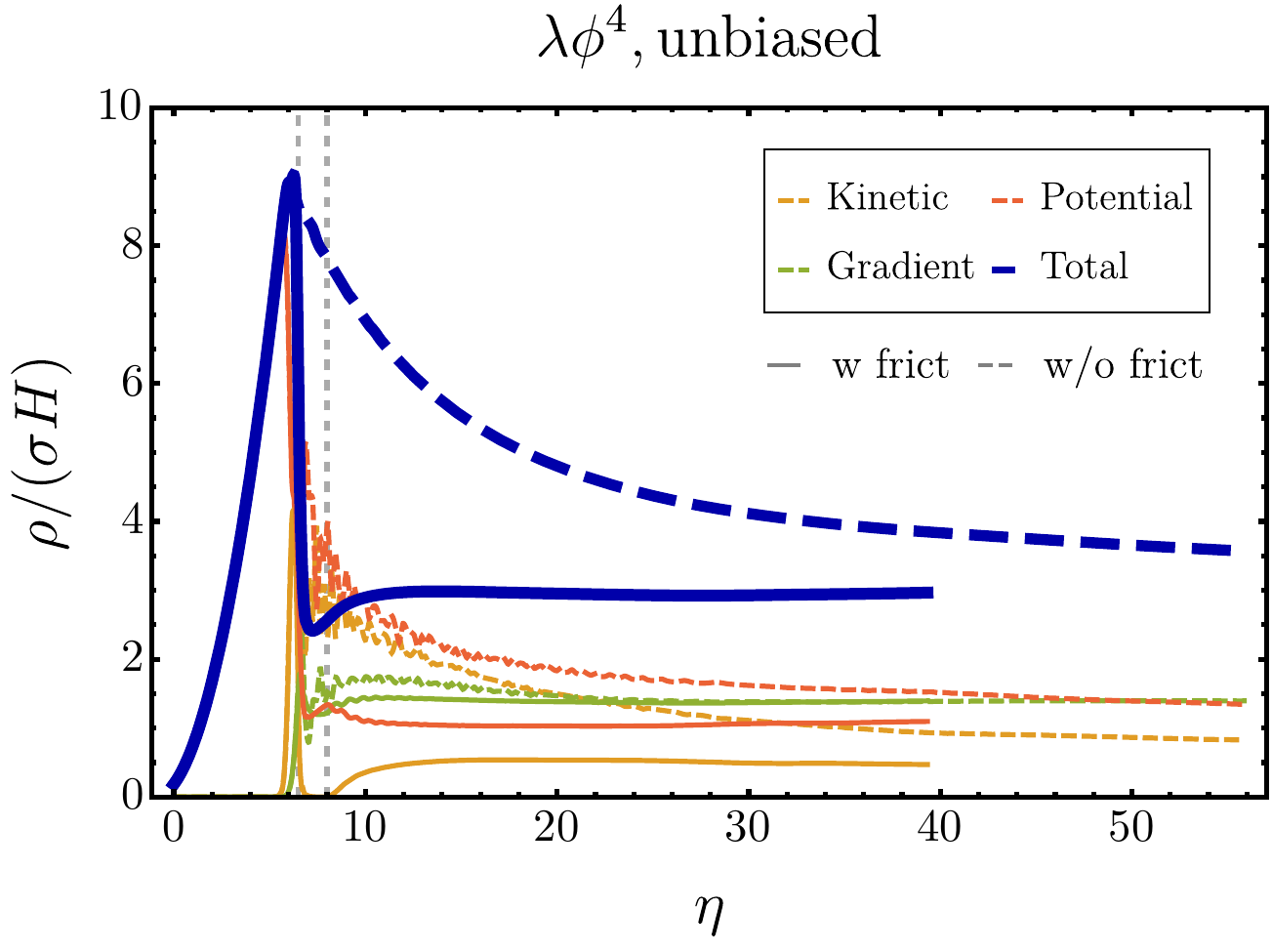}
    \end{center}
    \caption{\textit{Left:} Evolution of the area parameter for two simulations of the unbiased $\phi^4$ scenario, one including the friction term \eqref{eq:fricteq} (solid curve) and without it (dashed curve). The lattice parameters of the simulations are $(N,L) = (3060,70)$ and $(N,L)=(4960,70)$ respectively, and both use second-order spatial derivatives. \textit{Right:} Evolution of the energy density and its components for the same two simulations. The vertical dashed lines in both panels delimit the times when friction is active. }
    \label{fig:frict-comp}
\end{figure*}

\subsection{Friction} \label{app:friction}

In our simulations, we have added a friction term $+a^2 (\eta) \Gamma (a) \phi'$ on the left hand side of Eq.~\eqref{eq:eomscalar} in order to accelerate the achievement of the scaling regime by the DW network, with $\Gamma(a)$ a function depending on the scale factor as follows,
\be \Gamma (a) \equiv \frac{\Gamma_0/m}{(1 + e^{\gamma (a (\eta_1) - a)})(1 + e^{\gamma (a - a (\eta_2))} )} \ , \label{eq:fricteq}\ee
where $\Gamma_0$ and $\gamma$ are constants, and $\eta_1$ and $\eta_2$ delimit the time interval when friction is active.
All the results presented in this work, both for unbiased and biased potentials, have been obtained with parameters $\eta_1 = 6.5$, $\eta_2 = 8$, $\Gamma_0 = 0.343$, and $\gamma = 50$. This choice allows to maximize the freezing of the DWs during a short period of time.

In order to confirm the validity of this approach, in Fig.~\ref{fig:frict-comp} we compare the evolution of the area parameter and energy density components between two simulations of the scalar field dynamics, one with friction term (with $N=3060$ and $L=70$) and one without it (with $N=4980$ and $L=70$), for the unbiased quartic potential. In order to trust our results, we must ensure that the dynamics of the domain walls at late times (i.e.~well within the scaling regime) are unchanged by the friction term. The area parameter without friction (left panel of Fig.~\ref{fig:frict-comp}) reaches the expected scaling value $\mathcal{A} \approx 1$ at $\eta \approx 20$, while with friction it reaches the same value earlier, at $\eta \approx 14$. Afterwards, the evolution of the area parameter is similar in both cases. Regarding the energy density (right panel), we observe that all its components achieve a local maximum at time $\eta \approx 6$ in both simulations. The onset of friction at $\eta_1 = 6.5$ makes the different energy components quickly settle towards their stable final values, which they attain at $\eta \approx 14$. In the absence of friction, the same phenomenon happens, but the stabilization takes much longer. The gradient component stabilizes around $\eta \approx 20$. However, the other two components (kinetic and potential) stabilize at much later times, and in fact, our lattice resolution (which allows to resolve the DWs until $\eta_{\rm max} \approx 60$) is not good enough to completely observe their convergence to the frictionless results. However, we have carried out simulations in 2+1-dimensions of the same system with $N^2=(48k)^2$ (using {\tt Clustereasy}~\cite{Felder:2007kat}), which allow to simulate the field until $\eta_{\rm max} \approx 200$, and observe that such stabilization also occurs for the kinetic and potential contributions, with their final values matching the results from simulations with friction.

\subsection{Fourth-order accurate spatial derivatives} \label{app:spatialderivs}

In order to solve the field dynamics in the lattice, we must replace the continuous spatial derivatives  in the equations of motion \ref{eq:eomscalar}-\eqref{eq:eomGW} by finite difference approximations. The standard version of \CL uses expressions accurate up to order $\mathcal{O}(\Delta x^2)$ in the lattice spacing. Most of the results of this paper have been obtained, however, with fourth-order accurate expressions. More specifically, following the recursive method of \cite{DiscreteDerivatives}, we use the following expressions for the first and second derivatives of a field $f$ in the $i$-spatial direction, i.e.~
\begin{align}
  [\nabla_i f]^{(4)}  \equiv & \frac{f_{{\bf n}-2 \hat{\imath}} - 8 f_{{\bf n}- \hat{\imath}} + 8 f_{{\bf n}+ \hat{\imath}}  - f_{{\bf n}+2 \hat{\imath}}}{12\, \Delta x}  \notag \\
    &\longrightarrow ~~ \partial_i{\tt f}({\bf x})\big|_{{\bf x}\,\equiv\, {\bf n}\dx} + \mathcal{O}(\Delta x^4) \ ,  \label{eq:der4a} 
\end{align}
and
\begin{align}
  [\nabla_i^2 f]^{(4)}  \equiv &\,\frac{- f_{{\bf n}+2  \hat{\imath}} + 16 f_{{\bf n}+ \hat{\imath}}  -30 f_{{\bf n} } + 16 f_{{\bf n}- \hat{\imath}} - f_{{\bf n}-2 \hat{\imath}}}{12\,\Delta x ^2}  \notag \\
    &\longrightarrow ~~ \partial_i^2{\tt f}({\bf x})\big|_{{\bf x}\,\equiv\, {\bf n} \Delta x} + \mathcal{O}(\Delta x^4) \label{eq:der4b} \ ,
\end{align}
where $f_{\bf x} \equiv f({\bf x})$, the vector ${\bf n} = (n_1, n_2, n_3)$ (with $n_i = 0, \dots N-1$) tags the lattice site where the derivative is computed, and $\hat{\imath}$ are unit vectors in the $i$-spatial direction (corresponding to positive displacements of length $\Delta x$). 

The use of fourth-order accurate derivatives allows to improve the accuracy of our results at intermediate and ultraviolet scales, in a less computationally-expensive way than e.g.~increasing the lattice number of points. In order to illustrate this, in Fig.~\ref{fig:GradientOrderComp} we compare the evolution of the GW spectrum for two simulations of the unbiased quartic scenario: one using the fourth-order expressions \eqref{eq:der4a}-\eqref{eq:der4b}, and another one using the standard second-order ones implemented in \CLns. We observe that, while the spectra at scales $x \equiv k /(2 \pi a H) \simeq 20$ coincide quite well, they show important differences at scales $x\geq 20$. The simulation with fourth-order accurate spatial derivatives reduces significantly the amplitude of the peak at very UV scales, confirming that it is a lattice artifact coming from lack of UV resolution in the lattice. Using fourth-order expressions also allows to better resolve the plateau at scales $x \sim 20 - 60$, formed after the achievement of the scaling regime.

\subsection{Spectrum dependence on lattice parameters} \label{app:GWspec-LNdep}

\begin{figure*}[!p]
    \centering

    \begin{minipage}[t]{0.48\textwidth}
        \centering
        \includegraphics[width=\textwidth]{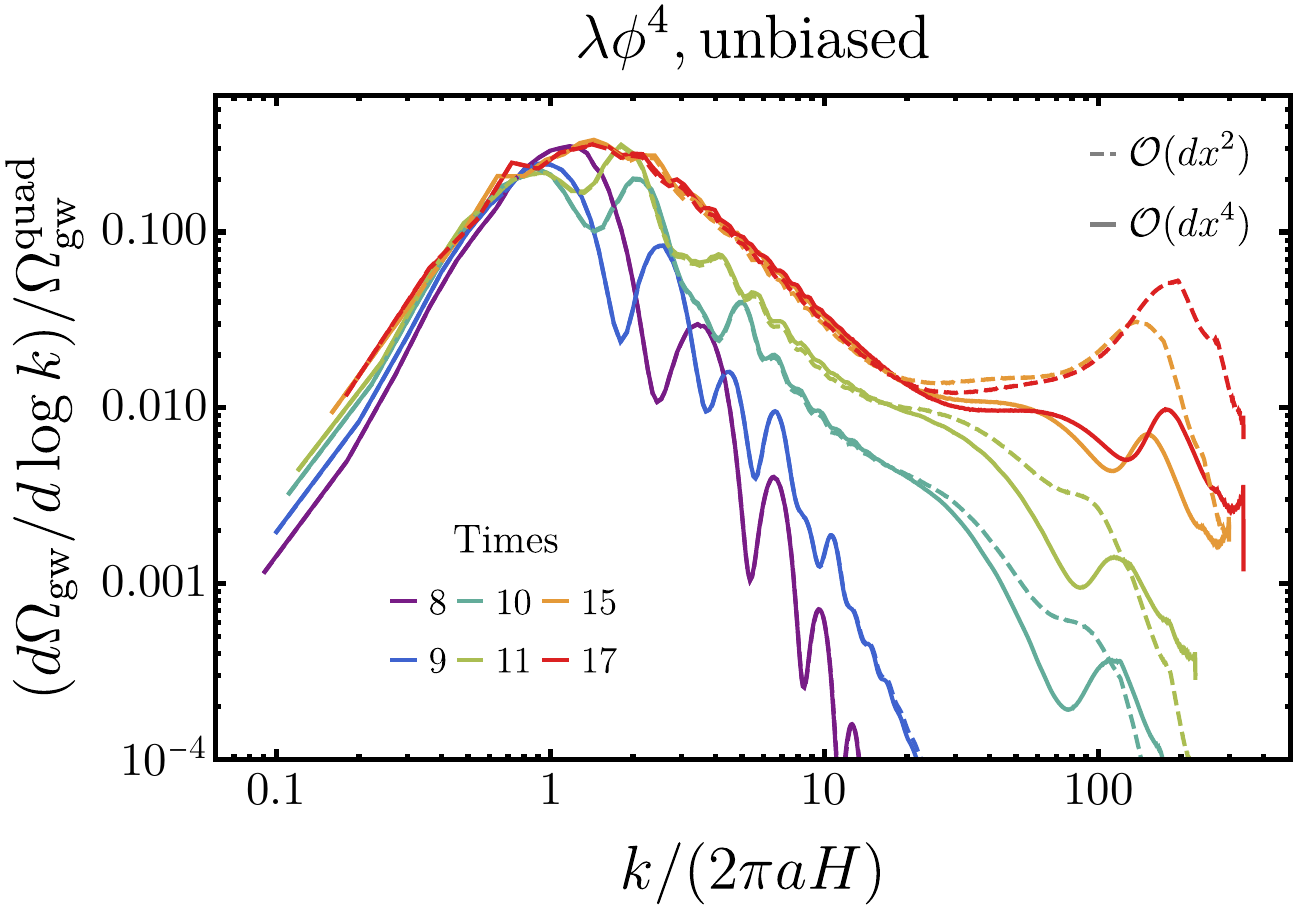} 
    \caption{Evolution of the GW spectrum for the unbiased $\phi^4$ potential with $N=3060$ and $L=141$. The solid lines use the fourth-order accurate expressions \eqref{eq:der4a}-\eqref{eq:der4b}, while the dashed lines use the second-order ones implemented in the standard version of \CLns.}
    \label{fig:GradientOrderComp}
    \end{minipage}
    \hfill
    \begin{minipage}[t]{0.48\textwidth}
    \centering
    \includegraphics[width=\textwidth]{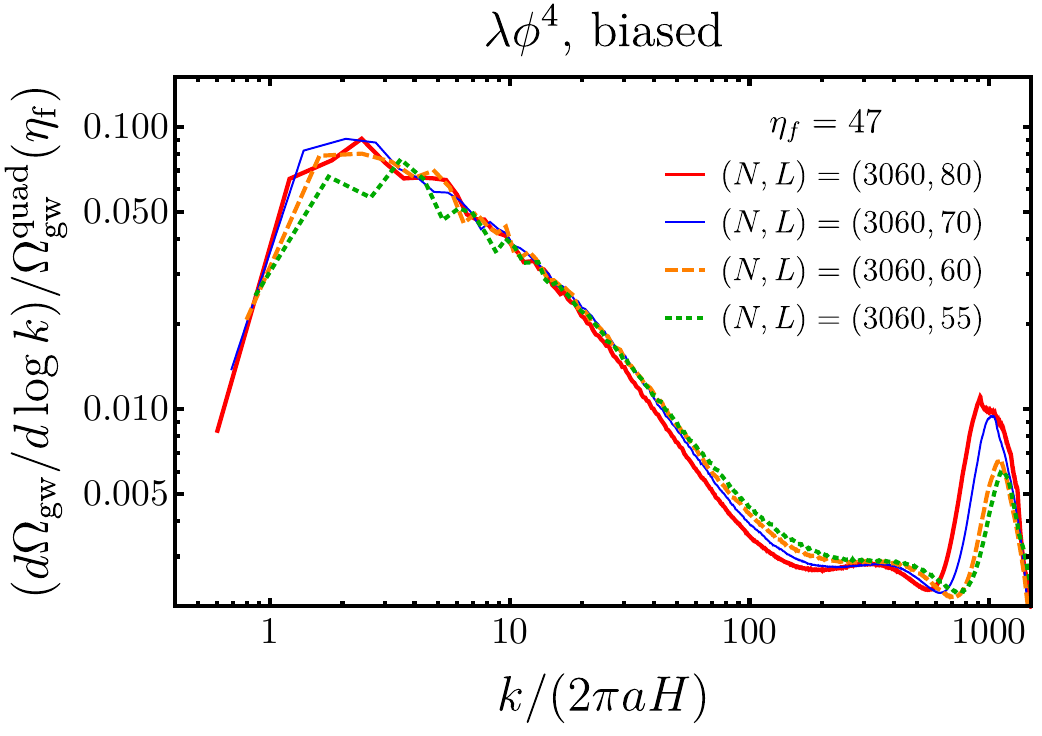} 
    \caption{Comparison of the GW spectrum for the biased $\phi^4$ potential at the final time $\eta_f = 47$, for $N=3060$ and different values of $L$.}
    \label{fig:LNComp}
    \end{minipage}

    \vspace{1.5cm}

    \begin{minipage}[t]{0.48\textwidth}
        \centering
        \includegraphics[width=\textwidth]{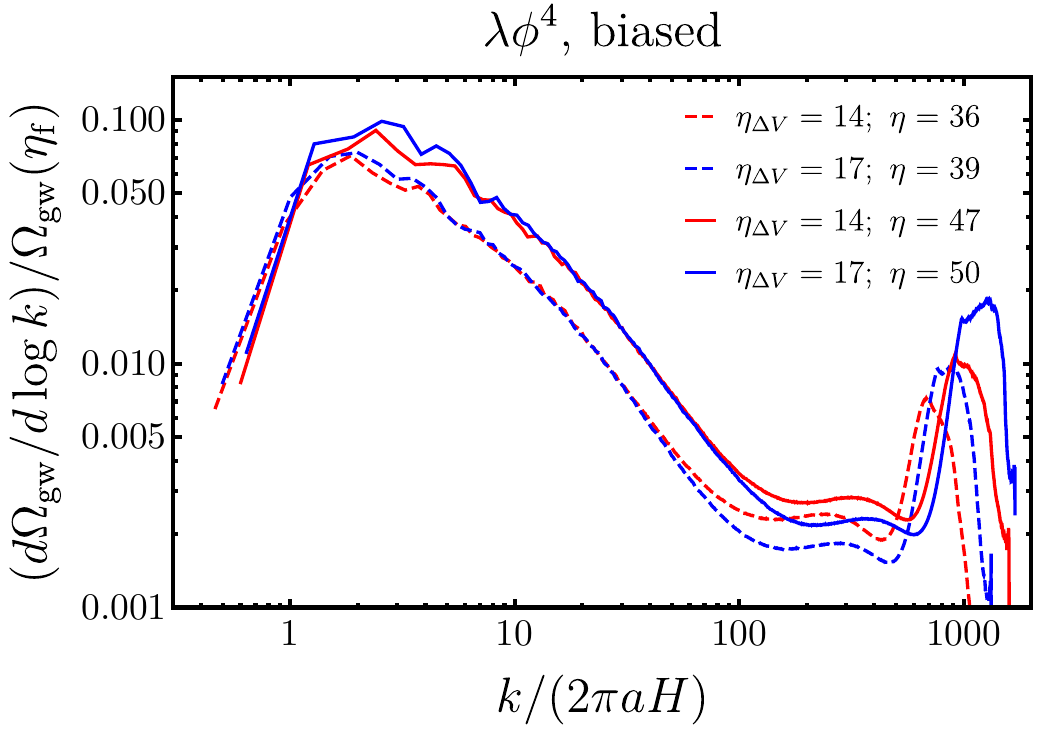}
        \caption{Comparison of GW spectra with $\eta_{\Delta V} = 14$ and $\eta_{\Delta V} = 17$, both at an intermediate time (dashed) and the final one (solid). Both simulations are obtained for lattice parameters $N=3060$, $L=80$.}
        \label{fig:Bias17}
    \end{minipage}
    \hfill
    \begin{minipage}[t]{0.48\textwidth}
        \centering
        \includegraphics[width=\textwidth]{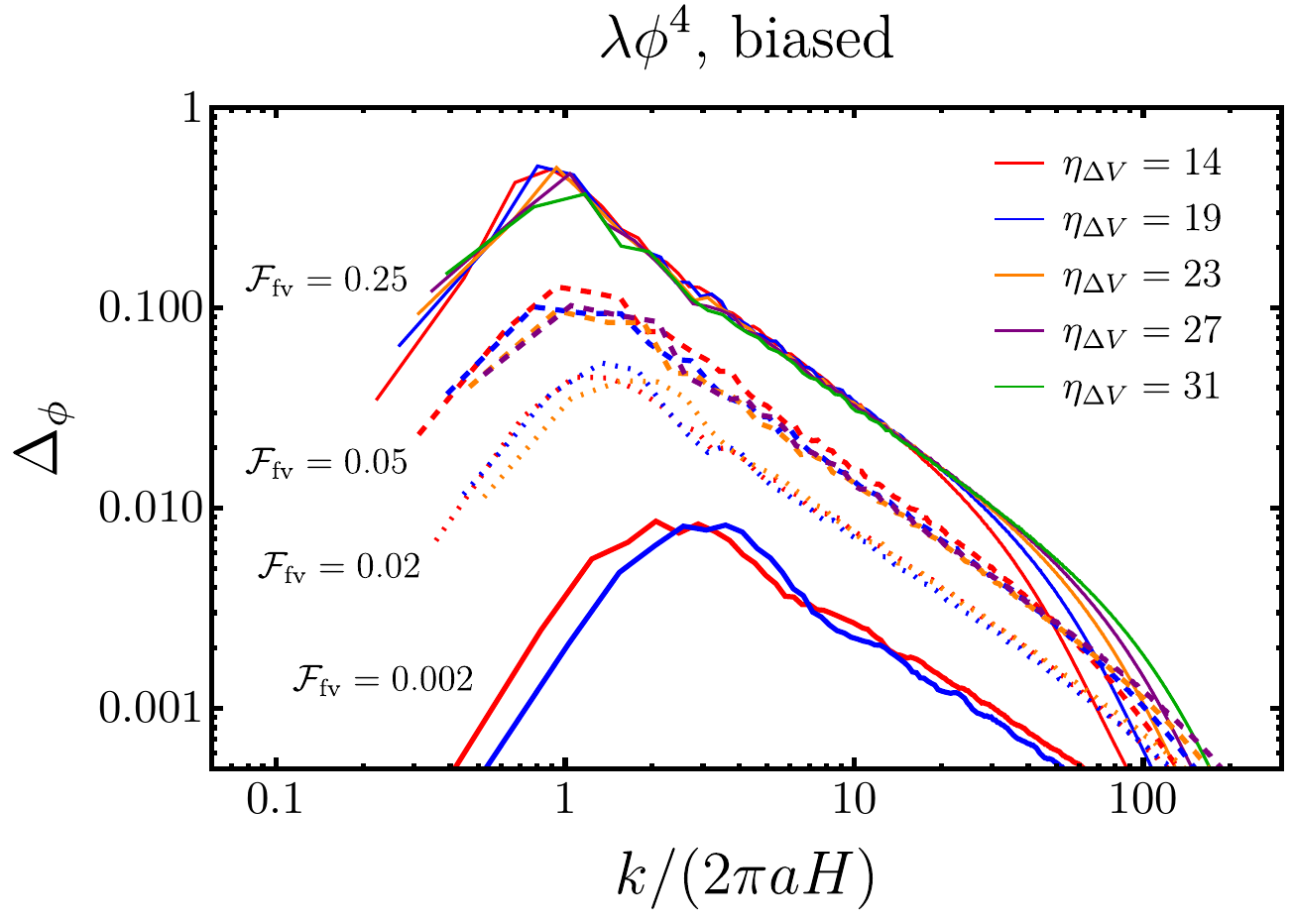}
        \caption{ Scalar field spectra for different bias sizes (represented by different coloured lines) extracted from simulations with $N=5000$, $L=90$, and second-order accurate spatial derivatives. The spectra are shown when the false vacuum fraction is $\mathcal{F}_{\rm fv}=0.25$ (solid), 0.05 (dashed), 0.02 (dotted), and 0.002 (solid thick). Note that for the largest $\eta_{\Delta V}$, we only show the spectra for the largest values of $\mathcal{F}_{\rm fv}$, since smaller values cannot be attained within the range of validity of our simulations.}
        \label{fig:ScalarSpectraEtaDV}
    \end{minipage}
\end{figure*}

Given a lattice of $N$ points per dimension and side length $L$, the domain walls can be well resolved approximately until time $\eta_{\max}$, defined in Eq.~\eqref{eq:tmax}. Our results have mainly been obtained with simulations in lattices of $N=3060$ and $L=80$, which corresponds to $\eta_{\max} \approx 32$. At this time, most of the domain walls have decayed, and the dominant source of GWs are scalar waves. However, as shown in the snapshots of Fig.~\ref{fig:snapshots}, some domain walls still survive at $\eta \approx 36$, and completely annihilate shortly thereafter. In order to validate the robustness of our results, in Fig.~\ref{fig:LNComp} we compare the GW spectrum at the final time of our simulation, $\eta_f = 47$, obtained with our benchmark lattice parameters ($N = 3060$, $L = 80$), to three additional simulations with the same $N$ but smaller side lengths, $L = 70,\,60,\,55$, corresponding to $\eta_{\max} = 37, 43,\,47$ respectively. The amplitude and shape of the spectrum are consistent across all three simulations, with the peak being better resolved for larger $L$ as expected. However, the amplitude of the spectrum at wavenumbers $x = k/(2 \pi a H) \geq 50$ shows a slight dependence on $L$ due to the changing UV lattice resolution. For this reason, we exclude this region from the fits presented in the main text.

\clearpage

\subsection{Spectrum dependence on bias size} \label{app:BiasSize}

In the simulations presented in Sec.~\ref{sec:biased} we have fixed the size of the bias term (the coefficient $q$) such that $\eta_{\Delta V} =14$. In order to assess the robustness of our GW spectrum parametrization under changes of the bias size, in Fig.~\ref{fig:Bias17} we compare the final GW spectrum obtained with one realization of the case $\eta_{\Delta V} = 14$, with another one for the slightly larger value $\eta_{\Delta V} = 17$. We observe that both spectra are remarkably similar in the region $x<50$.

Ideally, we would like to simulate the gravitational waves for larger $\eta_{\Delta V}$, in order to further confirm the robustness of our spectrum parametrization. However, this delays the annihilation of the domain walls and hence requires larger lattices, beyond our available computational resources. Alternatively, we have been able to simulate the scalar field only (without gravitational waves) in lattices of up to $N=5000$ points per dimension, which successfully captures the annihilation process for smaller bias sizes. In Fig.~\ref{fig:ScalarSpectraEtaDV} we compare the scalar field spectrum, i.e.~$\Delta_\phi \equiv k^3 \lvert\phi_k\lvert^2/(2\pi^2) $, for different bias sizes in the range $\eta_{\Delta V} \in [14, 31]$, at different times during the annihilation. More specifically, we compare the spectra when the false vacuum fraction (i.e.~the fraction of the lattice volume where $\phi>0$) attains the values $\mathcal{F}_{\rm fv} = 0.25, 0.05, 0.02, 0.002$. We can observe that the shapes of the scalar spectrum at equal values of $\mathcal{F}_{\rm fv}$ are similar for different values of $\eta_{\Delta V}$, and since the scalar field sources the GW spectrum, we expect the latter to also be roughly independent of the bias size.

\subsection{Comparison with previous results} \label{sec:SpectraComp}

\begin{figure}[t]
        \centering
        \includegraphics[width=0.48\textwidth]{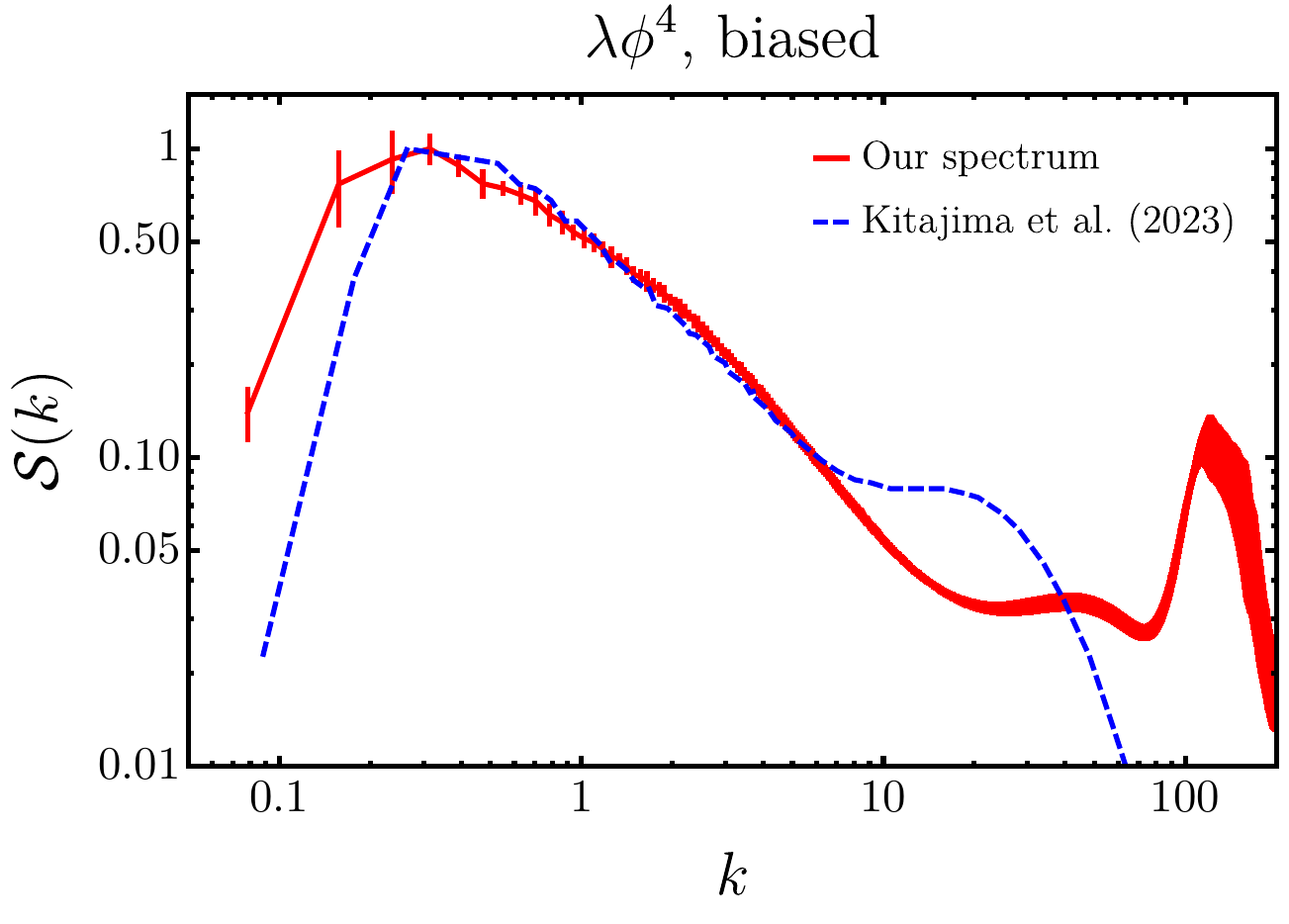}
        \caption{Comparison of our result for the GW spectrum in biased $\phi^4$ case (red) with the one obtained in Ref.~\cite{Kitajima:2023cek} (dashed blue), in terms of comoving momentum and normalized to unity at the peak.}
        \label{fig:Comparison}
\end{figure}

In Fig.~\ref{fig:Comparison} we compare our prediction for the shape of the GW spectrum in the biased $\phi^4$ case with a time-independent bias, with the one obtained in Ref.~\cite{Kitajima:2023cek} for a time-dependent one (see the top panel of Fig.~3 of that work). Both spectra agree quite well in the position of the peak and the slope of the UV tail in the region $0.3 \lesssim k \lesssim 5$. On the other hand, the IR tail observed in \cite{Kitajima:2023cek} is steeper than ours. Both spectra also show a plateau at high frequencies, with a difference of a factor two in its position.

\newpage

  \bibliography{References.bib,extra.bib}
  \bibliographystyle{utphys}

\end{document}